\documentclass[10pt]{iopart}
\usepackage{graphicx}
\usepackage{bm}
\usepackage{bbold}
\expandafter\let\csname equation*\endcsname\relax
\expandafter\let\csname endequation*\endcsname\relax
\usepackage[utf8]{inputenc}
\usepackage{amsmath}
\usepackage{amssymb}
\usepackage{amsfonts}
\usepackage{mathtools}
\usepackage{graphicx}
\usepackage{fixme}
\usepackage{color}
\usepackage{braket}
\usepackage[font=small]{caption}
\usepackage{cite}

\usepackage{url, hyperref}

\definecolor{myred}{RGB}{214,26,70}
\definecolor{myreddark}{RGB}{76,8,38}
\definecolor{myblue}{RGB}{35,106,185}
\definecolor{mybluedark}{RGB}{19,56,99}
\definecolor{mybluebright}{RGB}{225,236,249}

\def\te{{\rm e}}

\def\bk{{\bf k}}
\def\bp{{\bf p}}

\def\calO{\mathcal{O}}

\def\Tr{{\rm Tr}}
\def\TrB{{\rm Tr }_{\rm B }}
\def\T{{\mathcal T}}

\def\pa{\partial}
\def\nn{\nonumber}

\def\mB{ {m_{\rm B} } } 
\def\TB{ {\mathcal{T}_{\rm B} } }
\def\tB{{t_{\text{B}}}}
\def\aB{{a_{\text{B}}}}
\def\nB{{n_{\text{B}}}}
\def\rhoB { { \rho_{\rm B } } }
\def\rhoI { { \rho_{\rm I } } }

\def\Re{{ \rm Re }}
\def\Im{{ \rm Im }}
\def\tp{{ \tilde{p} }}
\def\tk{{ \tilde{k} }}

\def\BEC{{ \rm BEC }}

\begin{document}
\title{Critical slowdown of non-equilibrium polaron dynamics}
\date{\today}

\author{K. \ Knakkergaard \ Nielsen$^1$, L. \ A. \ Pe{$\tilde{\rm n}$}a Ardila$^1$, G. \ M. \ Bruun$^1$, T. \ Pohl$^1$}
\address{$^1$ Department of Physics and Astronomy, Aarhus University, Ny Munkegade, DK-8000 Aarhus C, Denmark}

\ead{kristianknakkergaard@phys.au.dk}

\begin{abstract} 
We study the quantum dynamics of a single impurity following its sudden immersion into a Bose-Einstein condensate. The ensuing formation of the Bose polaron in this general setting can be seen as impurity decoherence driven by the condensate, which we describe within a master equation approach. We derive rigorous analytical results for this decoherence dynamics, and thereby reveal distinct stages of its evolution from a universal stretched exponential initial relaxation to the final approach to equilibrium. The associated polaron formation time exhibits a strong dependence on the impurity speed and is found to undergo a critical slowdown around the speed of sound of the condensate. This rich non-equilibrium behaviour of quantum impurities is of direct relevance to recent cold atom experiments, in which Bose polarons are created by a sudden quench of the impurity-bath interaction. \\

\end{abstract}

\maketitle

\section{Introduction} \label{sec.introduction}
Understanding the non-equilibrium dynamics of many-body systems remains an outstanding challenge in physics. Cold atomic gases have emerged as an excellent platform to explore this question \cite{Altman2015}, since they can be well isolated from their environment and offer an extraordinary level of control. In particular, the precise tunability of interactions through Feshbach resonances \cite{Grimm2010} has opened the door to studies of interaction effects in quantum many-body systems with unprecedented control. While this offers unique perspectives for studying paradigmatic models in condensed matter physics, the scope of cold atom research has extended well beyond such initial ideas. Exciting new research directions include the dynamical emergence of thermal equilibrium in isolated quantum systems \cite{Altman2015}, and the observation of many-body localization \cite{Bloch2017}, linked to the breakdown of ergodicity \cite{Anderson1958, Anderson1980}. Another example is the polaron quasiparticle, which was originally introduced by Landau \cite{Landau1933} to describe the interaction of electrons with the atomic crystal of a solid, and has since been employed to understand a broad range of problems in condensed matter physics \cite{Mahan2000book}. Experiments on imbalanced Fermi gases~\cite{Schirotzek2009,Grimm2012,Kohl2012,Demler2016b,Roati2017} provide an ideal quantum simulation platform for the Fermi polaron \cite{Mora2010,Massignan2014,schmidt18}. 
At the same time, the possibility to realize so-called Bose polarons in atomic Bose-Einstein condensates (BECs) \cite{Arlt2016, Jin2016} has raised further questions and ushered in new theoretical investigations \cite{Devreese2009, Schmidt2013, Sarma2014, Casteels2014, Levinsen2015, Giorgini2015, Georg2015, Giorgini2016,Palma2017, Grusdt2017, Guardian2018, Guardian2018b,Yoshida2018}, expanding our understanding of quantum impurity physics. The properties of the Bose polaron is arguably closer to the generic solid-state polaron, since the surrounding BEC has a linear low energy dispersion in analogy with acoustic phonons in a solid. While most of previous efforts have been directed towards the equilibrium properties of the Bose polaron, its dynamics has spawned theoretical work only recently \cite{Demler2016,Demler2014,Grusdt2017.2,Lausch2017,Lewenstein2018,Ashida2018}, predicting the formation of phonon-impurity bound states for strongly interacting systems \cite{Demler2016} and studying trajectories and momentum relaxation of moving impurities \cite{Demler2014,Grusdt2017.2,Lausch2017,Lewenstein2018}, as well as the dynamics of phonon dressing in spinor condensates \cite{Ashida2018}.

\begin{figure}[h!]
\begin{center}
\includegraphics[width=1.0\columnwidth]{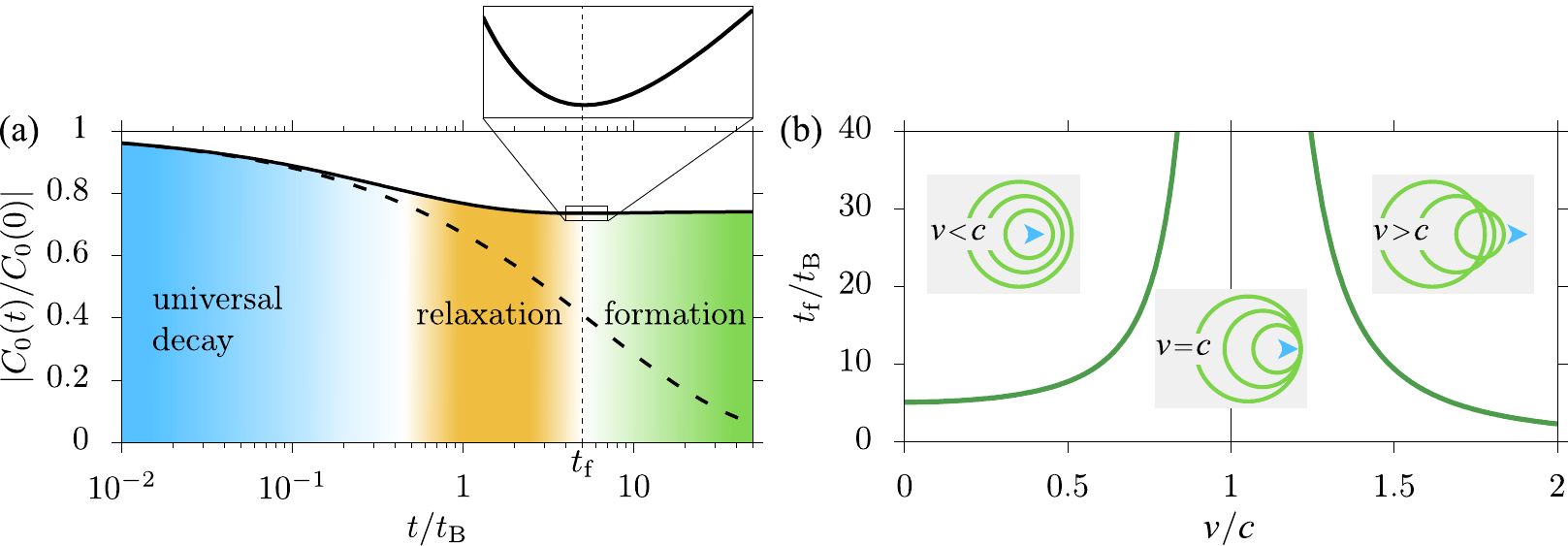}
\caption{(a) Non-equilibrium dynamics of the coherence of a quantum impurity immersed into a Bose-Einstein condensate at time $t=0$. The two timescales $\tB$ and $t_{\rm f}$ define three distinct evolution stages that characterize the approach of the coherence $|C_p|$ towards its equilibrium value. The solid line illustrates this dynamics for a boson-boson scattering length of $\aB/\xi=0.01$ and an impurity at rest with a boson-impurity scattering length $a/\xi=0.1$ in the equal mass case $\nu = m / \mB = 1$. The dashed line shows the corresponding dynamics for a non-interacting BEC in which the universal initial decay $|C_p(t)|=|C_p(0)|\exp(-\sqrt{t/t_0})$, set by a third characteristic timescale $t_0$, persists throughout the entire time evolution of the quantum impurity. The inset shows that the formation time $t_{\rm f}$ is defined as the instant where the coherence is minimal. (b) The speed dependence of the polaron formation time $t_{\rm f}$ reveals a critical slowdown as the impurity speed approaches the speed of sound $c$. The insets illustrate the emission of sound waves by the moving impurity and its inhibition around the speed of sound.}
\label{fig.coherence_layout}
\end{center}
\end{figure}

While the quantum evolution of the Bose polaron has briefly been studied \cite{Demler2016}, in the present paper we show that the formation dynamics of the impurity towards the Bose polaron undergoes a critical slowdown for finite impurity momenta. This is made possible by a master equation approach from which we derive rigorous results that reveal distinct evolution stages, as illustrated in figure \ref{fig.coherence_layout}(a). Within this framework, the Bose polaron emerges as a consequence of impurity-bath decoherence, whereby all many-body states apart from the polaron ground state dephase as the phonon dressed impurity approaches its steady state. In this picture, the critical slowdown manifests itself as arrested decoherence, whereby phonon dissipation into the environment is rendered inefficient as the impurity speed approaches the speed of sound (see figure \ref{fig.coherence_layout}(b)). While analytical results are derived for weak impurity interactions, much of the identified characteristic features of the non-equilibrium impurity dynamics turn out to be of greater generality and occur for arbitrary coupling strengths.

\section{The System} \label{sec.system}
We consider an impurity of mass $m$ in a gas of bosons with a mass $\mB$ and a density $\nB$. The boson-boson and boson-impurity interactions are both of short range nature and characterized by the scattering lengths $\aB$ and $a$, respectively. For weak interactions and close to zero temperature, the bosons form a BEC that is accurately described by Bogoliubov theory. The Hamiltonian, 
\begin{equation}\label{eq.Hamiltonian}
H = H_{\rm B} + H_{\rm I} + H_{\rm IB},
\end{equation}
of this system can be separated into three terms, where
\begin{equation}\label{eq.HB}
H_{\rm B} = \sum_{\bk} E_\bk \beta^\dagger_{\bk} \beta_{\bk}
\end{equation}
describes the BEC in terms of Bogoliubov modes with momenta $\bk$ and associated energies $E_\bk = [\varepsilon^{\rm B}_\bk (\varepsilon^{\rm B}_\bk + 2\nB\TB)]^{1/2}$ that are created by the operators $\beta_{\bk}^\dagger$. Here, $\TB=4\pi\aB/\mB$ is the scattering matrix for the boson-boson interaction, and $\varepsilon^{\rm B}_\bk = k^2 / 2\mB$ is the bare boson kinetic energy at momentum $\bk$. We work in units where $\hbar = 1$ and the temperature is zero. The impurity Hamiltonian is given by
\begin{equation}\label{eq.HI}
H_{\rm I} = \sum_{\bk}\varepsilon_\bk c^\dagger_{\bk}c_{\bk},
\end{equation}
where $c^\dagger_{\bk}$ creates an impurity with momentum $\bk$ and energy 
\begin{equation}\label{eq.ep}
\varepsilon_\bk = k^2 / 2m + \varepsilon_{\rm MF}, 
\end{equation}
which includes the mean-field shift $\varepsilon_{\rm MF}=\nB U_0$ of the impurity energy due to its interaction with the BEC. 
Here, $U_\bk$ is the impurity-boson interaction in momentum space. Accordingly, the term 
\begin{equation} \label{eq.HIB}
H_{\rm IB}= \sum_{\bk, \bp}U_\bk \sqrt{\frac{\nB \varepsilon^{\rm B}_\bk}{VE_\bk}}c^\dagger_{\bp - \bk}c_{\bp}\left(\beta^\dagger_{\bk} + \beta_{-\bk} \right)
\end{equation}
describes the impurity-boson interaction. Equations \eqref{eq.HB}-\eqref{eq.HIB} correspond to the so-called Fr\"ohlich model \cite{Frohlich1952}, originally put forth to describe the electrons coupled to optical phonons of a dielectric crystal. In the present case of an impurity in a BEC, there are interaction terms not included in \eqref{eq.HIB}, which describe the scattering of the impurity on bosons already excited out of the BEC. Their contribution is, however, suppressed by a factor $(\nB a_{\rm B}^3)^{1/2}$ \cite{Georg2015}, which we assume to be small. Focussing our analysis on the regime of validity of the Fr\"ohlich Hamiltonian, we consider all observables to second order in the impurity scattering length $a$. We note however, that central characteristics of the impurity dynamics, such as the stretched exponential initial coherence decay and the critical slowdown of polaron formation, shown in figure \ref{fig.coherence_layout}, persist beyond the Fr\"ohlich model.

Explicitly, we express the interaction as $U_\bk = U_0 g_\bk$, with $g_\bk$ a rescaled coupling and $g_0 = 1$. We then solve the Lippmann-Schwinger equation to express $U_0$ in terms of $a$ to second order. This yields $U_0 = \T + \T^2 / (2\pi)^3 \int {\rm d}^3 k \; g_\bk^2 \; 2m_r / k^2$. Here $\T = 2\pi a / m_r$ is the zero energy impurity-boson scattering matrix and $m_r = \mB m /(m + \mB)$ denotes the reduced mass. This allows to re-express our results in terms of the scattering length $a$, which yields well-defined result in the limit of zero range interactions.
 
In order to study decoherence, we consider a quench in which an impurity with momentum $\bp$ is suddenly immersed into the condensate at time $t = 0$, creating the initial state
\begin{equation}
\ket{\psi_0} = \left(\cos\theta + \sin\theta c_{\bp}^\dagger \right) \ket{\BEC},
\label{eq.initialstate}
\end{equation}
where $\theta$ is the mixing angle that determines the probability, $\sin^2\theta$, for initial impurity creation. The coherence between the vacuum and the single-impurity state can then be obtained from 
\begin{align}
C_p(t)=\bra{\psi_0}\! c_{\bp}(t) \!\ket{\psi_0},
\label{eq.coherence0}
\end{align} 
where $c_{\bp}(t)=\exp(iHt)c_{\bp}\exp(-iHt)$ is the time-dependent impurity operator in the Heisenberg picture. 
The evolution of $C_p(t)$ closely traces the dynamical formation of the polaron. Substituting the initial state \eqref{eq.initialstate} into \eqref{eq.coherence0} yields 
\begin{align}
C_p(t) = &\cos\theta \sin\theta \! \bra{\BEC} \! c_{\bp}(t)c^\dagger_{\bp}(0) \! \ket{\BEC} = i \cos\theta \sin\theta G_{\bp}(t).
\label{eq.linktoGreensfunction}
\end{align} 
This shows that the coherence is related to the time-dependent impurity Green's function, $G_\bp$, for $t > 0$. For long times, it should thus approach the asymptotic Green's function of the polaron
 \begin{align}
\lim_{t\rightarrow\infty}C_p(t) \sim Z_pe^{-iE_p\cdot t-t/2\tau_p},
\label{eq.Longtimelimit}
\end{align} 
where $E_p$ is the polaron energy, $\tau_p$ its lifetime, and $Z_p$ its quasiparticle residue. For an infinite polaron lifetime, $|C_p(t)|$ therefore converges to the 
quasiparticle weight $Z_p$ \cite{Demler2016}.

In cold atom experiments, the initial state \eqref{eq.initialstate} can be prepared by driving a transition between internal atomic states of the impurity which feature different interactions. Indeed recent experimental demonstrations of the Bose polaron were based on RF spectroscopy on a hyperfine transition between a weakly and a strongly interacting state of a single-component $^{39}$K BEC \cite{Arlt2016} or of $^{40}$K impurity atoms immersed in a $^{87}$Rb condensate \cite{Jin2016}. In both cases, a direct measurement of the coherence dynamics is possible via Ramsey spectroscopy \cite{Cetina2015,Cetina2016,Widera2018}.

\section{Master equation description} \label{sec.masterequation}
We determine the dynamics of the coherence using the impurity density operator $\rhoI$. First, we consider the density operator $\rho(t)$ of the entire system, i.e. the impurity and the BEC. In the interaction picture, it obeys the von Neumann equation 
\begin{equation}
i\pa_t\rho = [H_{\rm IB}(t), \rho(t)].
\label{eq.vonNeumanninitial}
\end{equation}
One can formally solve this equation by integrating both sides and resubstituting the result into the right hand side of \eqref{eq.vonNeumanninitial} to obtain
\begin{equation}\label{eq.dtrho}
\pa_t\rho = \!-i[H_{\rm IB}(t), \rho(0)] - \!\int_0^t \!\!{\rm d}s [H_{\rm IB}(t), [H_{\rm IB}(s), \rho(s)]]. \!\!
\end{equation}
We can now trace out the bosonic degrees of freedom on both sides of \eqref{eq.dtrho} to obtain an evolution equation for the reduced density operator, $\rhoI = \TrB\rho$, of the impurity. For our chosen initial state \eqref{eq.initialstate}, the initial density operator of the entire system factorizes according to $\rho(0) = \rhoI(0) \otimes \rhoB(0)$, where $\rhoB(0) = \ket{\BEC}\bra{\BEC}$ is the density operator for the BEC. Taking the trace of the first term in \eqref{eq.dtrho} only yields terms proportional to $\braket{\beta} = \braket{\beta^\dagger} = 0$, such that we obtain the following equation of motion for the impurity density operator 
\begin{equation}
\pa_t\rhoI =\! -\!\int_0^t \!\!{\rm d}s \;\TrB\!\left[H_{\rm IB}(t),[H_{\rm IB}(s), \rhoI(s) \otimes \rhoB(0)] \right]. \!
\label{eq.rhoIequation}
\end{equation}
Here, we have made the Born approximation, $\rho(s) = \rhoI(s) \otimes \rhoB(0)$ assuming that the density matrix of the BEC is unaffected by the impurity, which is justified for small impurity interactions. In the same limit, we can also replace $\rhoI(s)$ by $\rhoI(t)$ to obtain a time-local equation that contains all relevant contributions up to second order in the interaction strength \cite{BreuerPetruccione, Petruccione2004}. Altogether this makes it possible to evaluate the trace over the commutator in \eqref{eq.rhoIequation}. Using further that $\bra{\BEC}\beta_{\bk}\beta^\dagger_{\bk^\prime}\ket{\BEC} = \delta_{\bk , \bk^\prime}$ and $\beta_{\bk}(t) = \te^{-iE_\bk t}\beta_{\bk}(0)$ in 
the interaction picture, we finally obtain 
\begin{align}
\pa_t \rhoI =& -\frac{\nB \T^2}{V} \! \sum_{\bk, \bp 1, \bp 2} \!\! g_\bk^2 \frac{\varepsilon^{\rm B}_\bk}{E_\bk}\int_0^t {\rm d}s \left(e^{-i E_\bk(t-s)} \right. \nn \\
&\times \left. \left[c_{\bp 1 - \bk}^\dagger(t) c_{\bp 1}(t), c_{\bp 2 + \bk}^\dagger(s) c_{\bp 2}(s)\rhoI(t)\right] + \text{h.c.}\right).
\label{eq.rhoIequationfinal}
\end{align}
Further details of this derivation are given in \ref{app1}. Equation \eqref{eq.rhoIequationfinal} constitutes an effective von Neumann equation for the impurity density operator, and will now be used to determine the dynamics of the impurity coherence.

\section{Impurity Coherence} \label{sec.secondordercoherence}
Knowing the time-dependent density operator of the impurity we can calculate its coherence from
\begin{align}
C_p(t) &= \Tr [c_{\bp}(t) \rhoI(t)] = \bra{0}c_{\bp}(t)\rhoI(t)\ket{0},
\label{eq.Coherence}
\end{align}
where we now trace over the impurity degrees of freedom, and have used that only the vacuum $\ket{0}$ contributes to the trace for the initial state \eqref{eq.initialstate}. It turns out to be convenient to introduce the coherence $\tilde{C}_p(t)=\te^{i\varepsilon_\bp t}C_p(t)$, stripped of the single particle and mean field phase rotation. Using \eqref{eq.Coherence} and $c_{\bp}(t) = \te^{-i\varepsilon_\bp t}c_{\bp}(0)$ we can write for its time evolution $\pa_t\tilde C_p(t)=\bra{0}c_{\bp}(0)\pa_t\rhoI(t)\ket{0}$, which, upon substituting \eqref{eq.rhoIequationfinal} and after some algebra, yields 
\begin{align}
\pa_t \tilde C_p(t) = i[\Gamma_p(t) - \Gamma_p(0)]\tilde C_p(t),
\label{eq.Diff}
\end{align}
with the time-dependent rate coefficient
\begin{align}
\Gamma_p(t) &= \nB\mathcal T^2 \int \frac{{\rm d}^3 k}{(2\pi)^3} g^2_\bk \frac{\varepsilon^{\rm B}_{\bk}}{E_{\bk}}\frac{\te^{i(\varepsilon_{\bp} - \varepsilon_{\bp - \bk} - E_{\bk})t}}{\varepsilon_{\bp} - \varepsilon_{\bp - \bk} - E_{\bk}}.
\label{eq.Gamma}
\end{align}
Equation \eqref{eq.Diff} implies a pure decay of the impurity coherence driven by its interaction with the surrounding BEC. The solution of \eqref{eq.Diff} is readily obtained, and upon reintroducing the phase factor $\te^{i\varepsilon_\bp t}$ gives
\begin{equation}
C_p(t)=\te^{-i(p^2/2m + \nB\T + \Sigma_p)t}\te^{i\int_0^t\!{\rm d}s\Gamma_p(s)}C_p(0),
\label{eq.MainResult}
\end{equation}
with 
\begin{equation}
\hspace{-0.02cm}\Sigma_p \!= \nB \T^2 \!\!\int \!\! \frac{{\rm d}^3 k}{(2\pi)^3} g^2_\bk \!\left[\frac{\varepsilon^{\rm B}_\bk}{E_\bk}\frac{1}{\varepsilon_\bp - \varepsilon_{\bp - \bk} - E_\bk} \! + \!\frac{2m_r}{k^2}\right]. \!\!
\label{eq.Energy}
\end{equation}
Here, we have used 
\begin{align}
\varepsilon_{\bp} + \Gamma_p(0) &= p^2 / 2m + \nB U_0 + \Gamma_p(0) = p^2 / 2m + \nB \T + \Sigma_p
\end{align}
which follows from the Lippmann-Schwinger equation, relating $U_0$ to the zero-energy impurity-boson scattering matrix $\T$, as discussed in Sec. \ref{sec.system}. Equation \eqref{eq.MainResult} explicitly shows how the phase factor corresponding to the polaron energy $E_p= p^2 / 2m + \nB \T + \Sigma_p$ naturally emerges from our formalism. Indeed, \eqref{eq.Energy} coincides with the second order contribution to the ground state energy of the Bose polaron obtained in \cite{Casteels2014,Georg2015} in the limit of a zero range potential with $g_\bk=1$. Comparing \eqref{eq.MainResult} with the $t\rightarrow\infty$ limit given by \eqref{eq.Longtimelimit}, we see that the integral $\int_0^t\!{\rm d}s \;\Gamma_p(s)$ determines the dynamical formation of the polaron state. 

Upon evaluating the integral in \eqref{eq.Gamma}, the expression for $\Gamma_p(t)$ can be written as 
\begin{equation}
\Gamma_p(t)=\Gamma\cdot\gamma(\nu, p\xi/\sqrt{2}, t/t_{\rm B}),
\end{equation}
where 
\begin{equation}
\Gamma=\frac{m_{\rm B}^2}{m_r^2} \frac{\sqrt{2}a^2}{\aB\xi}t_{\rm B}^{-1},
\label{eq.rate_constant}
\end{equation}
is a rate constant, and $\gamma(\nu, p\xi/\sqrt{2}, t/t_{\rm B})$ is a time-dependent function that only depends on the mass ratio $\nu=m/\mB$ and the dimensionless impurity momentum $p\xi/\sqrt{2}$ in units of the inverse of the BEC coherence length $\xi=1 /\sqrt{8\pi\nB\aB}$. Moreover, \eqref{eq.rate_constant} contains the characteristic timescale $\tB=\xi/\sqrt{2}c$ for the impurity dynamics, determined by the speed of sound $c = \sqrt{4\pi\nB\aB} / \mB$ of the condensate. Physically, this timescale reflects how fast Bogoliubov modes can build up distortions on a length scale $\xi$, which corresponds to the typical size of the screening cloud surrounding the impurity.

\section{Short-time dynamics} \label{sec.short_time_dynamics}
At very short times, the impurity dynamics depends on the actual shape of the impurity-boson interaction. To model a finite range, we consider a step function in momentum space, $g_\bk = \Theta(\kappa - k)$. This choice provides a simplified model for the actual interaction between the atoms, whose typical range, $r_{0}$, relates to the characteristic momentum $\kappa\sim r_{0}^{-1}$. As shown in \ref{app2}, finite range effects are significant only at times shorter than $t_i / \tB \simeq 1 / (\kappa\xi)^2 \sim (r_{0}/\xi)^2$. The characteristic extent of interactions between alkaline atoms is below $1$nm and, thus, about two orders of magnitudes smaller than the typical coherence length $\xi\sim0.1\mu$m of atomic BECs. It thus follows that the finite interaction range only affects the initial dynamics for extremely short times $t / \tB \lesssim 10^{-4}$, such that we can take the well defined zero-range limit, $\kappa\rightarrow\infty$, under typical experimental conditions.

\begin{figure}[t!]
\begin{center}
\includegraphics[width=1.0\columnwidth]{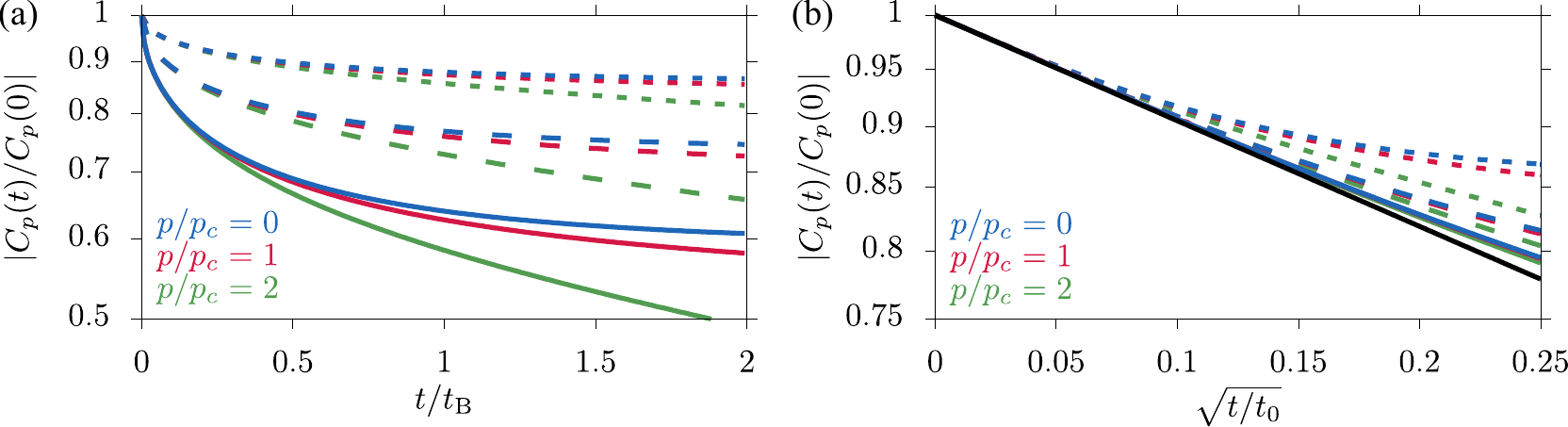}
\caption{Coherence dynamics for different indicated impurity momenta, $p$, and for different impurity-boson scattering lengths, $a / \xi = 0.07$ (dotted lines), $a / \xi = 0.1$ (dashed lines) and $a / \xi = 0.13$ (solid lines). For $t\ll \tB$, all of the different curves shown in panel (a) collapse to a single universal decay law $|C_p|\sim \exp(-\sqrt{t/t_0})$ [black line in (b)] as demonstrated in panel (b). The remaining parameters are $\nu = m / \mB = 1$ and $\aB/\xi = 0.01$.}
\label{fig.short_time}
\end{center}
\end{figure}

Figure \ref{fig.short_time}(a) shows the resulting short-time dynamics for different impurity momenta, $p$, and different values of the scattering length $a$. The initial behaviour appears to proceed independently of the impurity momentum, which can be readily understood from \eqref{eq.Gamma}. At early times $t$, it takes large momenta $k$ for the phase to rotate significantly in the exponential factor in the integrand for $\Gamma_p(t)$. The dominant contributions to the integral therefore stems from large momenta that eventually exceed the impurity momentum and render the initial impurity decoherence independent of $p$. By expanding the integral for $\Gamma_p(t)$ in \eqref{eq.Gamma} for short times, $t \ll \tB$, one finds (see \ref{app3}) 
\begin{equation}
C_p(t) = C_p(0)\te^{ -i\left(p^2/2m + \nB \T\right) t }\, \te^{-(1 + i)\sqrt{t / t_0}}.
\label{eq.coherence_short_time}
\end{equation}
The coherence thus undergoes a decay described by a stretched exponential with a timescale 
\begin{align}
t_0 = \frac{m_r}{ 16\pi n_{\rm B}^2 a^4},
\label{eq.coherence_short_timescale}
\end{align}
that only depends on the scattering length $a$ and the condensate density, while it is independent of $p$ and $\aB$. 
This is illustrated in figure \ref{fig.short_time}(b), where we show $|C_p(t)/C_p(0)|$ for the same momenta and scattering lengths as in figure \ref{fig.short_time}(a), but as a function of $\sqrt{t/t_0}$ instead of $t/\tB$. All data points indeed collapse to a single curve given by \eqref{eq.coherence_short_time} for times $t\ll t_{\rm B}$. 
The initial coherence dynamics thus undergoes a stretched exponential decay $|C_p(t)| = |C_p(0)|\exp(- \sqrt{t / t_0})$, prior to establishing the eventual polaron steady state as discussed in the next section. 

It is interesting to note that the discussed short-time dynamics reflects the asymptotic form of the impurity spectral function at large frequencies \cite{Demler2014,Braaten}. The latter has been pointed out to exhibit a universal form \cite{Braaten,Wild,Demler2014} involving the two-body contact of the condensate \cite{Braaten,Tan}.

\section{Long-time dynamics} \label{sec.longtime_dynamics}
\begin{figure}[t!]
\begin{center}
\includegraphics[width=0.5\columnwidth]{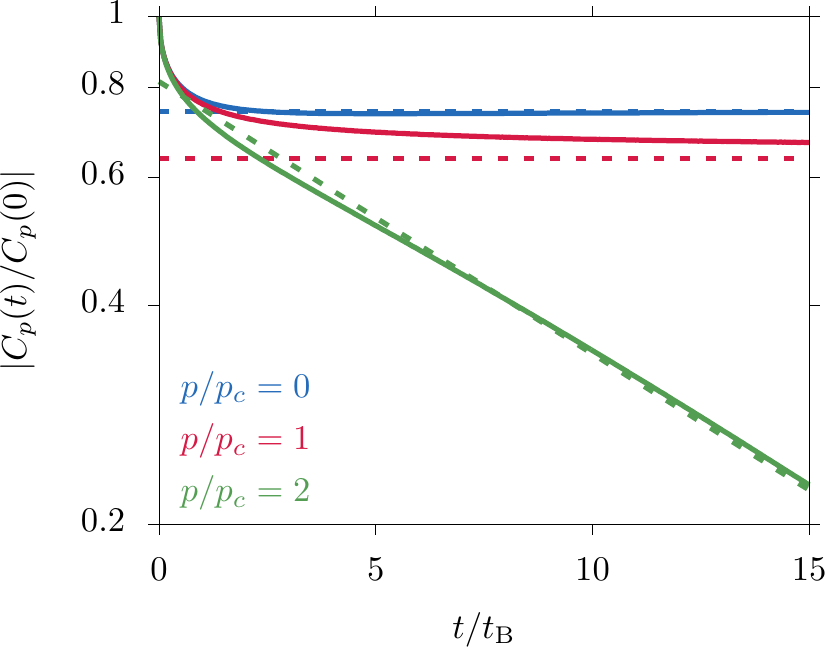}
\caption{Coherence dynamics for different indicated impurity momenta $p$. The solid lines show the derived solution of the master equation, which approaches the expected polaron steady state shown by the dashed lines. Specifically, below $p_c$ the coherence approaches the quasiparticle residue $Z_p$ given in \eqref{eq.residue}, while above $p_c$ it acquires a nonzero damping rate $1/(2\tau_p)$ given in \eqref{eq.GoldenRule}. The depicted behaviour indicates that the final approach towards this steady state becomes very slow for $p / p_c \approx 1$. The remaining parameters are $\nu = m / \mB = 1$, $\aB/\xi = 0.01$ and $a/\xi \!=\!0.1$.}
\label{fig.Coherence_long_times}
\end{center}
\end{figure}

During the subsequent evolution stage, $t \gtrsim t_{\rm B}$, the impurity dynamics starts to depend on the momentum whereby $|C_p(t)|$ approaches a finite value for $p<p_c$ but continues to decay above the Landau critical momentum $p_c=m c$ (see figure \ref{fig.Coherence_long_times}). 

\begin{figure}[t!]
\begin{center}
\includegraphics[width=0.5\columnwidth]{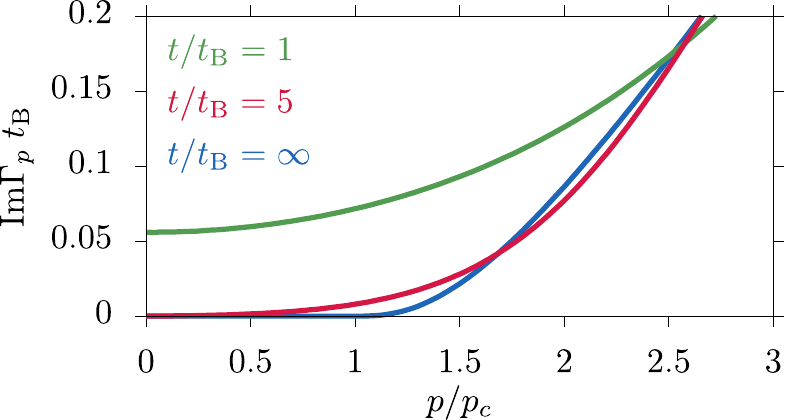}
\caption{Momentum dependence of the instantaneous damping rate $\Im\Gamma_p(t)$ at different indicated times in the impurity dynamics. 
For $p > p_c$ the impurity continues to decohere and therefore retains a finite decay rate, $1 / (2\tau_p) > 0$ given in \eqref{eq.GoldenRule}, as $t\rightarrow\infty$ (blue line). The remaining parameters are $\nu = m / \mB = 1$, $a/\xi = 0.1$ and $\aB/\xi = 0.01$.} 
\vspace{-0.6cm}
\label{fig.damping}
\end{center} 
\end{figure}

Indeed, by evaluating the time-dependent rate coefficient, \eqref{eq.Gamma}, in the limit of long times, we obtain the asymptotic decay time $\tau_p=1/[2{\rm Im}\Gamma_p(\infty)]$ (see \ref{app5})
\begin{equation}
\frac{1}{2\tau_p} = \nB \T^2 \int \frac{{\rm d}^3 k}{(2\pi)^3} \frac{\varepsilon^{\rm B}_\bk}{E_\bk} \pi \delta(\varepsilon_{\bp} - \varepsilon_{\bp - \bk} - E_{\bk}),
\label{eq.GoldenRule}
\end{equation}
which coincides with the Fermi golden rule expression for the spectral width of the polaron in equilibrium \cite{Casteels2014}. In figure \ref{fig.damping}, we show the instantaneous damping rate ${\rm Im}\Gamma_p(t)$. 
For $p<p_c$ the decay rate steadily decreases and vanishes at long times, reflecting the infinite lifetime of the eventually formed polaron since the impurity has insufficient kinetic energy to excite Bogoliubov modes in the condensate. For higher momenta, $p>p_c$, the impurity moves faster than the speed of sound and therefore emits Cherenkov radiation, causing decoherence and exponential decay on a timescale of $\tau_p$ given by \eqref{eq.GoldenRule}. We can separate this long-time behaviour in \eqref{eq.MainResult} by writing 
\begin{equation}
C_p(t) = \te^{-i E_p t- t/(2\tau_p)}\te^{\int_0^t{\rm d}s (i\Gamma_p(s) + 1 /(2\tau_p))}. \nn
\end{equation}
Since the integral over the rate coefficient becomes purely imaginary at long times, one can use this expression to define the asymptotic quasiparticle residue
\begin{equation}
Z_p = \te^{-\int_0^{\infty}{\rm d}s (\Im\Gamma_p(s) - 1 /(2\tau_p))}. 
\label{eq.residue}
\end{equation}
which coincides with the second order equilibrium result for the residue \cite{Georg2015} (see \ref{app6}). 

The real part of $\Gamma_p(t)$, however, always vanishes asymptotically such that the phase factor of $C_p(t)$ approaches $\exp(-iE_pt)$. For $p = 0$ and $\nu=1$, this long-time behaviour takes on a particularly simple form 
\begin{align}
\varepsilon_0 + \Sigma_0 - \Re \Gamma_0(t) =\nB \T + \Sigma_0\left(1 - 45 \frac{t_{\rm B} ^4}{ t^4}\right).
\label{eq.ImpurityEnergy}
\end{align} 
where $\Sigma_0=32\sqrt{2}\nB a^2/3m\xi$ can be obtained from \eqref{eq.Energy}. While a more general expression for arbitrary momenta and mass ratios is derived in \ref{app7}, this simple result shows that one can define an asymptotic impurity energy, which approaches the polaron ground state energy at long times. 

\section{Polaron formation time} \label{sec.formationtime}
Equation \eqref{eq.ImpurityEnergy} indicates that the timescale $t_{\rm f}$ for the discussed approach to equilibrium is on the order of $t_{\rm B}$ for equal masses $\nu=1$ and zero momentum. In general, however, the polaron formation time $t_{\rm f}$ can depend strongly on the impurity momentum, as we will now discuss.

\begin{figure}[!t]
\begin{center}
\includegraphics[width=1.0\columnwidth]{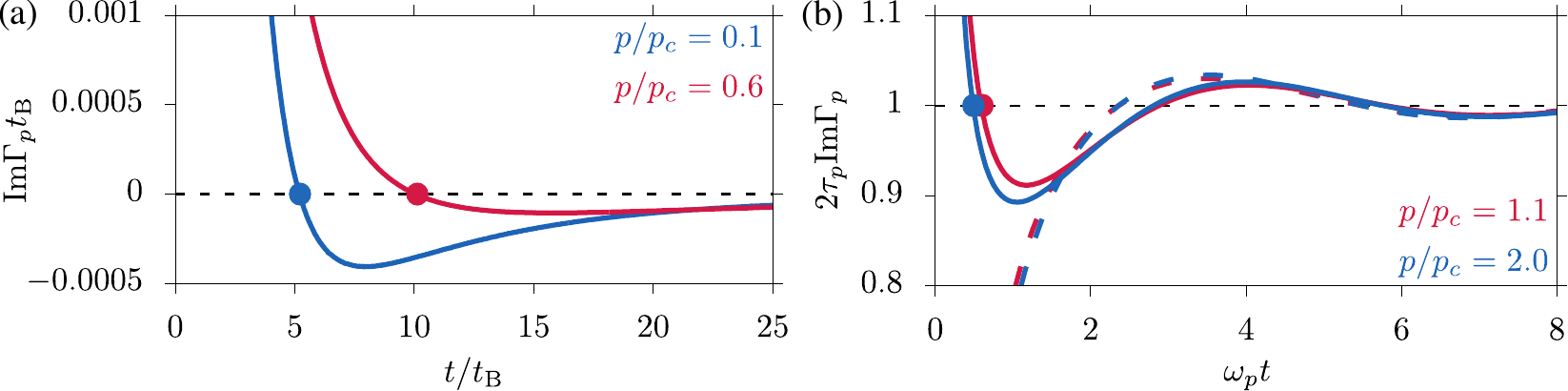}
\caption{Instantaneous damping rate $\Im\Gamma_p$ for different indicated impurity momenta $p$. The black dashed lines show the expected asymptotic value $1/(2\tau_p)$ given by \eqref{eq.GoldenRule}, which vanishes below the critical momentum [panel (a)] but is finite for $p>p_c$ [panel (b)]. The dots indicate when the damping rate crosses the long time limit, defining the formation time $t_{\rm f}$. The coloured dashed lines in panel (b) show the asymptotic rate \eqref{eq.ImGammap>pc_asymp}. The used mass ratio is $\nu = m / \mB = 1$. For panel (a) we use $a / \xi = 0.1$, $\aB / \xi = 0.01$.} 
\label{fig.ImGamma_asymp}
\end{center} 
\end{figure}

Consider first the regime of momenta below the Landau critical value, $p < p_c$. Here, we can gain insights into the asymptotic coherence dynamics by expanding \eqref{eq.Gamma} in orders of $1/t$. As outlined in \ref{app7}, this gives the following asymptotic behaviour
\begin{equation}
\ln\left|\frac{C_p(t)}{C_p(0)}\right| = - \frac{a^2}{\aB\xi}\left(A_p + B_p\frac{1}{(t/\tB)^2}\right).
\label{eq.cnormlongtimelimit}
\end{equation}
Both coefficients, $A_p=-(\aB \xi/a^2)\ln Z_p$ and $B_p$, depend only on the mass ratio, $\nu$, and the momentum ratio $p / p_c$. Regardless of these two parameters, though, one always finds that $B_p > 0$. It thus follows from \eqref{eq.cnormlongtimelimit} that $|C_p(t)|$ approaches its final value from \emph{below}. Since the coherence initially decreases, this implies that $|C_p(t)|$ always overshoots its eventual steady state and goes through a \emph{minimum} before starting its ultimate approach to the polaron steady state. The location of this minimum can therefore be used to define the polaron formation time, $t_{\rm f}$, which marks the onset of the final stage of the impurity dynamics, as illustrated in the inset of figure \ref{fig.coherence_layout}(a). Since $\pa_t |C_p(t)| = -\Im \Gamma_p(t)|C_p(t)|$, this is equivalent to finding the first zero of $\Im \Gamma_p$, as illustrated in figure \ref{fig.ImGamma_asymp}(a).

For larger impurity momenta $p > p_c$, the decoherence rate approaches a finite value $\Im \Gamma_p(\infty) = 1 /(2\tau_p)$ that reflects the emission of Cherenkov radiation as discussed in Sec. \ref{sec.longtime_dynamics}. 
We however find that the asymptotic decoherence rate approaches its steady state in an oscillatory fashion as 
\begin{equation}
\Im \Gamma_p(t) = \frac{1}{2\tau_p} \! + \! \frac{a^2}{\aB\xi}D_p(\nu) \frac{t_{\rm B}^{1/2}}{t^{3/2}}\cos\left(\omega_p t + \frac{\pi}{4}\right),
\label{eq.ImGammap>pc_asymp}
\end{equation}
where $D_p$ and $\omega_p$ only depend on $\nu$ and the scaled impurity momentum $p / p_c$. Equivalently to the low-momentum case above, we can thus determine the formation time $t_{\rm f}$ by $\Im \Gamma_p(t_{\rm f}) = 1 /(2\tau_p)$, defining the time when the decoherence rate first crosses its steady state value, as illustrated in figure \ref{fig.ImGamma_asymp}(b).

\begin{figure}[!t]
\begin{center}
\includegraphics[width=0.65\columnwidth]{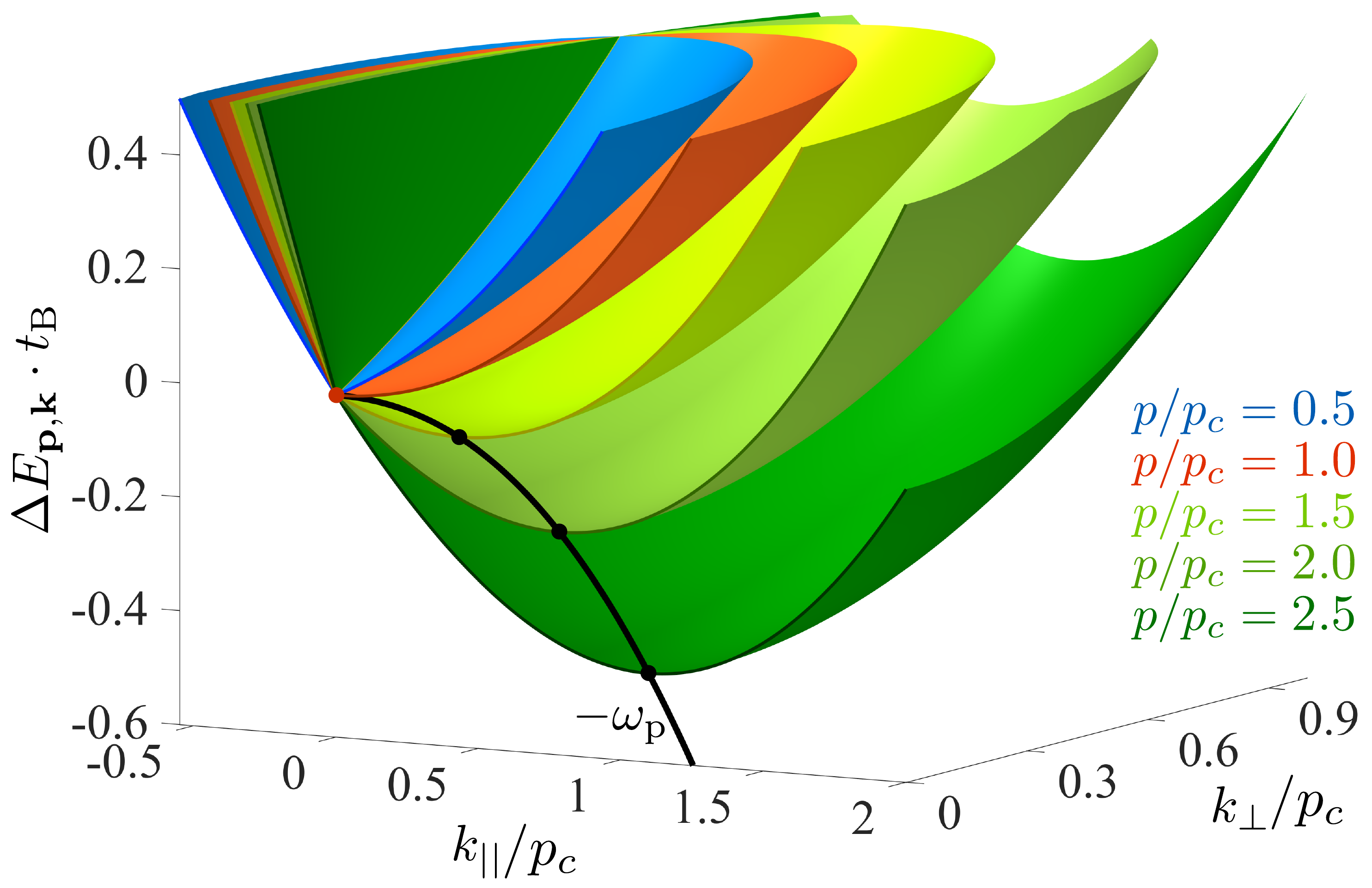}
\caption{Energy difference $\Delta E_{\bp,\bk} = \varepsilon_{\bp - {\bf k}} + E_{{\bf k}} - \varepsilon_{\bf p}$ associated with the emission of a phonon with momentum $\bk$ by an impurity moving with different indicated momenta $\bp$. The energy difference is shown as a function of the phonon momentum components parallel ($k_{||}$) and perpendicular ($k_\perp$) to the impurity momentum. The black dots indicate the minimum of $\Delta E_{\bp,\bk}$ which defines the frequency $\omega_p$ for $p > p_c$ according to Eqs. \eqref{eq.wp} and \eqref{eq.drwp}. As discussed in the text, the polaron formation time, $t_{\rm f}$, is related to the slope of $\Delta E_{\bp,\bk}$ around $k=0$, which vanishes for $p=p_c$ (red surface) and causes $t_{\rm f}$ to diverge. 
The used mass ratio is $\nu = m / \mB = 1$.} 
\label{fig.Energy_difference}
\end{center} 
\end{figure}

As also shown in figure \ref{fig.ImGamma_asymp}(b), $\omega_p$ determines both the frequency as well as the damping of the oscillation, since scaling time by $\omega_p^{-1}$ renders the asymptotic coherence dynamics nearly independent of $p$. The derivation of \eqref{eq.ImGammap>pc_asymp}, given in \ref{app7}, shows that the frequency 
\begin{equation}\label{eq.wp}
\omega_p = -\Delta E_{\bp,\bk_0}=\varepsilon_{\bf p} - \varepsilon_{\bp - {\bf k}_0} - E_{{\bf k}_0},
\end{equation}
is given by the energy cost to emit a Bogoliubov excitation with momentum $\bk_0$ that lies at the stationary point 
\begin{equation}\label{eq.drwp}
\nabla_{\bk}\Delta E_{\bp,\bk}|_{\bk = \bk_0} = 0
\end{equation}
which minimizes $\Delta E_{\bp,\bk}$. This value stems from off-shell phonon emission with momenta around $\bk_0$, where the phases in \eqref{eq.Gamma} add up constructively. 
 As illustrated in figure \ref{fig.Energy_difference}, the phonon momentum $\bk_0$ is always parallel to the impurity momentum, making forward scattering the dominant mechanism for phonon generation. Such processes become less important at long times $t\rightarrow\infty$ when off-shell scattering is suppressed by energy conservation.

Our results for $t_{\rm f}$ are summarized in figure \ref{fig.coherence_layout}(b) for equal masses, $\nu = 1$. For $p = 0$, we see that $t_{\rm f}\sim\xi / c = \sqrt{2}\tB$, which is consistent with the developed picture of polaron formation in terms of a phonon-induced decoherence process. It implies that the formation time corresponds to the time it takes to propagate the presence of the impurity, i.e. the time to traverse the impurity screening cloud with a size $\xi$. 

Most strikingly, we find that the formation time diverges when the impurity momentum approaches the Landau critical momentum $p_c$ from below as well as from above. Right at $p = p_c$ the coherence dynamics undergoes a critical slowdown and never reaches the polaron steady state as illustrated in figure \ref{fig.Coherence_long_times}. Close to $p_c$ the formation time can be described by the following critical behaviour 
\begin{equation}
t_{\rm f} = b \frac{\tB}{|1 - p / p_c|^{\eta}},
\label{eq.t_formation_at_pc}
\end{equation}
where the proportionality constant $b$ depends on the mass ratio. For $\nu = 1$, the critical exponent is found to be $\eta=2$ above and below $p_c$. In fact, one can show from Eqs. \eqref{eq.wp} and \eqref{eq.drwp} that $\omega_p\tB = \nu (p / p_c - 1)^2 / 4$ close to $p_c$ (see \ref{app7}). Hence, there is a universal critical exponent of $\eta=2$ for $p > p_c$. For $p < p_c$, on the other hand, the exponent turns out to depend on the mass ratio $\nu = m/\mB$. Explicitly, we find that for a light impurity with $\nu \leq 1$ the critical exponent remains to be $\eta = 2$, while it acquires a mass dependence $\eta = 1 + (1 + 1 / \nu) / 2$ for $\nu \geq 1$. Note that the formation generally diverges as $\nu\rightarrow0$ (see \ref{app8}).

According to the above discussion, the discovered critical slowdown is linked to forward emission of phonons and can be understood as follows.
As suggested by the form of \eqref{eq.Gamma}, the formation time $t_{\rm f}$ is generally determined by how fast the described off-shell scattering processes with energy costs $\Delta E_{\bp,\bk}$ dephase, whereby the dominant contributions arise from energy differences around $\Delta E_{\bp,\bk}\approx0$. As illustrated in figure \ref{fig.Energy_difference}, for impurity momenta below $p_c$ this corresponds to long-wavelength excitations with wave vectors around $k\approx0$. As the impurity momentum approaches $p_c$, the linear slope of $\Delta E_{\bp,\bk}$ with respect to momentum of forward emitted phonons steadily decreases, such that the corresponding virtual scattering events continue to add in phase for longer and longer times. For $p=p_c$, the slope eventually vanishes, thus implying a divergence of the formation time, as seen in figure \ref{fig.coherence_layout}(b). 
Above $p_c$, $\Delta E_{\bp,\bk}$ also vanishes for finite phonon momenta, as shown in figure \ref{fig.Energy_difference}. This enables the emission of Cherenkov radiation with $\Delta E_{\bp,\bk}=0$ and leads to finite steady state damping with the rate \eqref{eq.GoldenRule}. The approach to this asymptotic behaviour, however, is again limited by the dephasing of virtual scattering events around $k\approx0$ and is, therefore, slowed down when the impurity momentum approaches $p_c$ from above in the same way as for $p<p_c$.

An intuitive picture for the resulting critical slowdown derives from the developed notion that the polaron arises from decoherence driven by the dissipation of phonons. Hereby the propagation of generated phonons away from the impurity transmits information about its presence and thereby diminishes the coherence $C_p$. This effective measurement and decoherence process, however, becomes inefficient as the impurity speed approaches the speed of sound $c = p_c/m$. In fact, when the impurity moves precisely with $c$, forward emitted phonons propagate at the same speed and cannot be dissipated away from the impurity (see figure \ref{fig.coherence_layout}(b)), such that decoherence becomes arrested. In this case, forward scattering of phonons turns into the bottleneck for impurity relaxation and ultimately causes a critical slowdown of polaron formation when approaching the Landau critical momentum $p_c = mc$.

\section{Conclusion} \label{sec.conclusion}
In summary, we have studied the non-equilibrium dynamics of a quantum impurity immersed into a Bose-Einstein condensate. By tracing the dynamics of the impurity coherence via a master equation approach we could identify three distinct dynamical regimes from a stretched exponential initial coherence decay and a subsequent phonon-driven relaxation to the final formation of the polaron steady state (figure \ref{fig.coherence_layout}(a)). While the underlying hierarchy of associated timescales was shown to hold for arbitrary impurity momenta, the ultimate formation of the polaron undergoes a critical slowdown when approaching the speed of sound of the condensate (figure \ref{fig.coherence_layout}(b)).

This established link between the equilibrium properties of the condensate and the dynamical critical behaviour of the non-equilibrium impurity state raises a number of open questions. For example it suggest that a similar slowdown might occur in condensates at finite temperature \cite{Gunther2018} and cause extraordinary long polaron formation times close to the critical temperature for Bose-Einstein condensation. Clarifying this question would indeed be important for future experiments on the Bose polaron at finite temperatures and across the BEC phase transition. 
For the parameters of a recent experiment \cite{Arlt2016}, in which Bose polarons have been created and studied by driving the $(F=1,M_F=-1)\rightarrow(F=1,m_F=0)$ hyperfine transition of a $^{39}$K BEC, we obtain for our relevant timescales $t_{\rm B}\approx0.2$ms, $t_0\approx14$ms and $t_{\rm f}\approx1$ms for $a=20\aB$ and $p=0$. It turns out that the shortest timescale, $t_{\rm B}=0.2$ms, is just about the duration of the microwave excitation pulse used in \cite{Arlt2016}, such that the predictions of the present work should be observable with present technology, enabling direct experimental access to the non-equilibrium properties of the Bose polaron.

The results of this work should also be relevant to other systems and similar problems, such as the decoherence of molecular rotational states in superfluid helium droplets \cite{Stapelfeldt2013}, whose equilibrium properties have been linked to quantum impurity physics in recent theoretical work \cite{Lemeshko2015,Lemeshko2017,Lemeshko2017b}. Following these ideas, the theoretical framework described in the present work appears also applicable for the description of the rotational dynamics of such molecules \cite{Stapelfeldt2013}. While the corresponding decoherence timescales are expected to be much shorter than for cold atom systems, they can be well resolved using fs-laser spectroscopy \cite{Stapelfeldt2013}.

The process of impurity decoherence is intrinsic to the formation of the polaron. Therefore, understanding its interplay with coherent driving of internal impurity states is also important for the quality of light matter interfaces involving polaron physics. Examples include the storage \cite{hofferberth2017} and propagation \cite{lukin2017} of quantum light in strongly interacting nonlinear media \cite{killian2018} or atomic BECs coupled to an optical cavity \cite{fleischhauer2016}.

While the derivation of analytical results and scaling relations of this work has been possible in the limit of weak impurity interactions, the central results of these derivations are of more general applicability.
For example, the found universal short-time behaviour $|C_p(t)/C_p(0)|\approx 1-\sqrt{t/t_0}$ applies to arbitrary strong impurity-boson interactions and remains valid beyond the Fr\"ohlich model. Moreover, the presented theory indicates that in the limit of an ideal Bose gas, the stretched exponential decoherence given by \eqref{eq.coherence_short_time} persists for all times of the impurity dynamics. Most importantly, the discovered divergence of the formation time for moving impurities around the Landau critical momentum could be traced back to the suppression of phonon emission when the impurity speed approaches the speed of sound. We note that this mechanism does neither rely on the presence of weak interactions nor on the specific simplified form of the Fr\"ohlich model, \eqref{eq.HIB}. While higher-order interaction effects can shift the precise value of the corresponding impurity momentum, e.g. through the renormalization of the polaron mass \cite{Demler2014}, the discovered critical slowdown of polaron formation should thus occur for any interaction strength including the regime of strong impurity coupling. Future cold atom experiments on the momentum-resolved impurity dynamics will make it possible to put these predictions to test.

\ack{
We thank Jan Arlt, Christopher Pethick, Matteo Zaccanti and Richard Schmidt for fruitful discussions. This work has been supported by the DNRF through a Niels Bohr Professorship.
} 

\newpage 

\appendix

\section{The effective von Neumann equation}\label{app1}
We derive the effective von Neumann equation, ~\eqref{eq.rhoIequationfinal}. We write the interaction Hamiltonian \eqref{eq.Hamiltonian} in the interaction picture and the equation of motion for the impurity density operator after tracing over the bath and making the Born-Markov approximation \eqref{eq.rhoIequation} 
\begin{align}
H_{\rm IB}(t) &= \sum_{\bk, \bp}U_\bk\sqrt{\frac{\nB \varepsilon^{\rm B}_\bk}{VE_\bk}}c^\dagger_{\bp - \bk}(t)c_{\bp}(t)\left(\beta^\dagger_{\bk}(t) + \beta_{-\bk}(t) \right), \nn \\
\pa_t\rhoI &= -\int_0^t {\rm d}s \TrB \; \left[H_{\rm IB}(t), [H_{\rm IB}(s), \rhoI(t) \otimes \rhoB(0)] \right]. \nn
\end{align}
Here $c_{\bp}(t) = \te^{-i\varepsilon_p t}c_{\bp}(0)$ and $\beta_{\bk}(t) = \te^{-iE_\bk t}\beta_{\bk}(0)$ in the interaction picture. We then let $I_i(t) = c_{\bp i-\bk i}^\dagger(t) c_{\bp i}(t)$ and $B_i(t) = \beta^\dagger_{\bk i}(t) + \beta_{-\bk i}(t)$ for $i = 1, 2$. Examining the commutator of the second line this yields terms like
\begin{align}
&\TrB [I_1(t) \otimes B_1(t), [I_2(s) \otimes B_2(s), \rhoI(t) \otimes \rhoB(0)]] \nn \\
&=\braket{B_1(t)B_2(s)} [I_1(t), I_2(s) \rhoI(t)] + \braket{B_2(s)B_1(t)} [\rhoI(t) I_2(s), I_1(t)], \nn
\end{align}
which can be calculated using only the identity $\Tr_{\rm B}[I \! \otimes \! B]\! = \! I \; \Tr(B)$ for an impurity operator $I$ and a bath operator $B$ when tracing out the bath part. The mean value is taken with respect to the BEC: $\braket{B} = {\rm Tr}[B \rhoB]$. At zero temperature the density operator for the bath, the BEC, is simply: $\rhoB = \ket{\BEC}\bra{\BEC}$. We then get $\braket{B_1(t)B_2(s)} =\braket{\beta_{-\bk 1}(t) \beta^\dagger_{\bk 2}(s)} = \te^{-iE_{\bk 1} t + iE_{\bk 2} s}\bra{\BEC} \beta_{-\bk 1} \beta^\dagger_{\bk 2} \ket{\BEC} = \te^{-iE_{\bk} (t - s)} \delta_{-\bk, \bk 2}$, with $\bk = \bk_1$. We then get
\begin{align}
\pa_t \rhoI =& -\frac{\nB}{V} \!\!\int_0^{t} {\rm d}s \!\! \sum_{\substack{\bk1, \bk2 \\ \bp1, \bp2}} \!\!U_{\bk1}U_{\bk2}\sqrt{\!\!\frac{ \varepsilon^{\rm B}_{\bk1} \varepsilon^{\rm B}_{\bk2} }{ E_{\bk1} E_{\bk2} } } \!\left( \Tr \beta_{-\bk 1}(t) \beta^\dagger_{\bk 2}(s) \right. \nn \\
&\times \left. \left[ c_{\bp 1 - \bk1}^\dagger(t)c_{\bp 1}(t), c_{\bp 2 - \bk 2}^\dagger(s)c_{\bp 2}(s) \rhoI(t) \right] + \text{h.c.} \right) \nn \\
=& -\frac{\nB \T^2}{V} \int_0^{t} {\rm d}s \sum_{\bk, \bp 1, \bp 2} \!\! g^2_{\bk}\frac{ \varepsilon^{\rm B}_\bk }{ E_\bk } \left( \te^{-iE_\bk(t - s)} \right. \nn \\
&\times \left. \left[ c_{\bp 1 - \bk}^\dagger(t)c_{\bp 1}(t), c_{\bp 2 + \bk}^\dagger(s)c_{\bp 2}(s) \rhoI(t) \right] + \text{h.c.} \right). \nn
\end{align}
In the second equality we write $U_\bk = U_0 g_\bk$ and replace $U_0$ by the impurity-boson scattering matrix at zero energy $\T = 2\pi a / m_r$, with $a$ the impurity-boson scattering length. This is consistent to second order in $a$. We hereby derived \eqref{eq.rhoIequationfinal}. 

\section{The role of the interaction range}\label{app2}
\begin{figure}[h]
\begin{center}
\includegraphics[width=0.6\columnwidth]{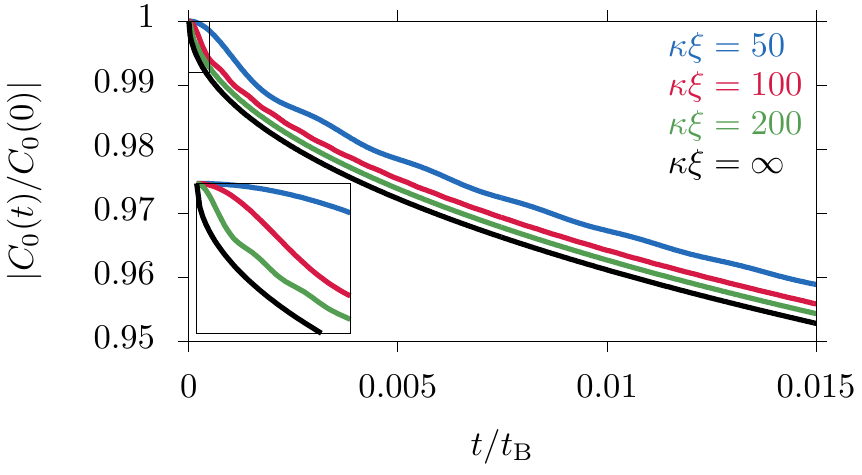}
\caption{Short-time behaviour of the coherence for different indicated values of momentum cut-off $\kappa$ that characterizes the range, $\sim 1/\kappa$, of the impurity-boson interaction potential. The inset illustrates the derived quadratic time dependence $\sim \exp(-(t/t_i)^2)$ at very short times $t\lesssim m_r/\kappa^2$, where $t_i$ scales as $t_i \sim 1 / \kappa^{3/2}$ with $\kappa$. Beyond this initial time, however, all curves approach the universal decay law $|C_p|\sim \exp(-\sqrt{t/t_0})$ as demonstrated in the main panel. The remaining parameters are $\nu = m / \mB = 1$, $a/\xi =0.1$ and $\aB/\xi = 0.01$.} 
\label{fig.Cnorm_momentum_cut}
\end{center}
\end{figure}

We here analyze the role of the interaction range as discussed briefly in Sec. \ref{sec.short_time_dynamics}. To model a finite range impurity-boson interaction, we consider a simple step function in momentum space, $g_\bk = \Theta(\kappa - k)$ where $\kappa$ is a high momentum cut-off. This choice provides a simplified model for the actual interaction between the atoms, whose typical range, $r_{0}$, relates to the characteristic momentum $\kappa\sim r_{0}^{-1}$. In figure \ref{fig.Cnorm_momentum_cut}, we plot the absolute value of the coherence for different values of the cut-off $\kappa$. As can be seen, the initial decoherence is characterized by a Gaussian damping $\ln |C_p(t)/C_p(0)|\sim -(t/t_\kappa)^2$. The characteristic time for this initial decoherence is determined by the cut-off as $t_\kappa \sim m_r / (\sqrt{\nB}a \kappa^{3/2})$. This simple initial behaviour applies for short times $t \lesssim t_i = m_r / \kappa^2$. Beyond this time, the decoherence proceeds in a way that is largely independent of $\kappa$.

\section{Short-time behaviour of $\Gamma_p(t)$}\label{app3}
We calculate the short-time behaviour of $\Gamma_p(t)$ for all momenta $p$ and mass ratios $\nu$. We further use this to calculate the coherence for short times and in the ideal Bose gas limit, $\aB \to 0$. \\

\begin{figure}[t!]
\begin{center}
\includegraphics[width=0.55\columnwidth]{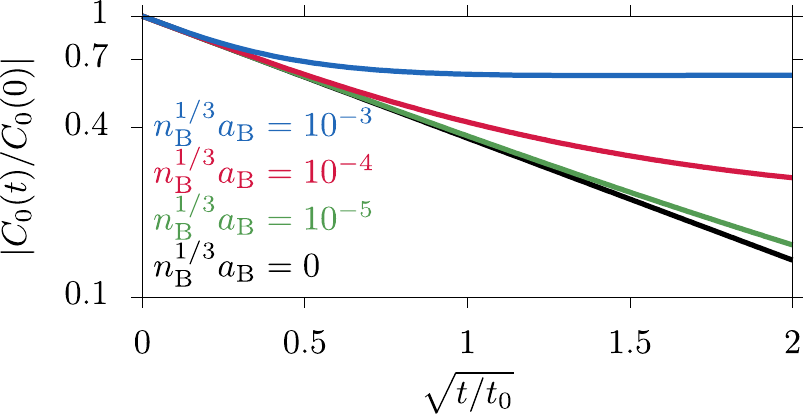}
\caption{The coherence is plotted as a function of $\sqrt{t / t_0}$ for different indicated values of $n_{\rm B}^{1/3}\aB$. For $\aB \to 0$ the decoherence dynamics of $|C_p(t)|$ collapse to the short-time behaviour $\propto \exp(-\sqrt{t / t_0})$ independent of the impurity momentum. The impurity-boson scattering length is $n_{\rm B}^{1/3}a = 0.1$. }
\label{fig.Cnorm_aB_depend}
\end{center}
\end{figure}

Using $\Phi_{\tp}(\tk) = -\tB(\varepsilon_\bp - \varepsilon_{\bp - \bk} - E_\bk) = \tk \sqrt{\tk^2 + 1} + (\tk^2 - 2 \tk \tp \cos \theta ) / \nu$ with $\tk = k\xi / \sqrt{2}$ and $\theta$ the angle between the impurity momentum $\bp$ and the phonon momentum $\bk$, we write \eqref{eq.Gamma} of the main text on unitless form 
\begin{equation}
\Gamma_p(t) = \! - \frac{\sqrt{2}(1 + 1 / \nu)^2}{4\pi \tB}\frac{a^2}{\aB \xi} \! \int {\rm d}\cos \theta {\rm d}\tk \frac{\tk^3}{\sqrt{1 + \tk^2}}\frac{\te^{-i\Phi_{\tp}(\tk)\tilde{t}}}{\Phi_{\tp}(\tk)}.
\label{eq.ImGamma2}
\end{equation}
Here $\cos\theta \in (-1, 1)$ and $\tk \in (0, \infty)$. We wish to expand \eqref{eq.ImGamma2} at short times $\tilde{t} = t / \tB \ll 1$. First, we infer a large momentum $\Lambda \gg \sqrt{2}/\xi$. We initially assume that we are at sufficiently short times so that for $\tk \leq \tilde{\Lambda} = \Lambda\xi / \sqrt{2}$: $\Phi_{\tp}(\tk) \tilde{t} \ll 1$. By expanding the exponential in the above integral we then get the following low momentum contribution
\begin{equation}
I^{\rm low}_p(t) = \int \cos\theta \int_0^{\tilde{\Lambda}}{\rm d}\tk \frac{\tk^3}{\sqrt{1 + \tk^2}}\left[\frac{1}{\Phi_{\tp}(\tk)} - i\tilde{t}\right].
\label{eq.low_momentum_contribution}
\end{equation}
It is clear that this integral contains contributions of $\calO(1)$ and $\calO(\tilde{t})$. Further, for $\tk > \tilde{\Lambda} \gg 1$ we can approximate the phase as $\Phi_{\tk}(\tp) = \tk\sqrt{\tk^2 + 1} + (\tk^2 - 2\tk\tp \cos\theta)/\nu \simeq \tk^2(1 + 1/\nu)$. Inserting this into \eqref{eq.ImGamma2} for $k > \Lambda$, we get the high momentum contribution
\begin{align}
I^{\rm high}(t) &= \int {\rm d}\cos \theta \int_{\tilde{\Lambda}}^{\infty} {\rm d}\tk \frac{\tk^3}{\sqrt{1 + \tk^2}}\frac{\te^{-i\Phi_{\tp}(\tk)\tilde{t}}}{\Phi_{\tp}(\tk)} \simeq \frac{2}{1 + 1 / \nu}\int_{\tilde{\Lambda}}^{\infty} {\rm d}\tk \; \te^{-i(1 + 1 / \nu)\tk^2 \tilde{t}}, \nn
\end{align}
which is independent of the impurity momentum, $p$. The integral is readily evaluated using the identity ${\rm erfi}(x) = -i{\rm erf}(i x) = 2/\sqrt{\pi} \; \int_0^x {\rm d}u \;\te^{u^2}$ for the so-called imaginary error function ${\rm erfi}$. Explicitly, we get
\begin{align}
I^{\rm high}(t) =& - \frac{1}{1 + 1 / \nu}\frac{1 + i}{\sqrt{2}}\sqrt{\frac{\pi}{(1 + 1 / \nu) \tilde{t}}} \left.{\rm erfi}\left(\frac{i - 1}{\sqrt{2}}\sqrt{(1 + 1 / \nu)\tilde{t}} \; \tk \right) \right|_{\tk = \tilde{\Lambda}}^{\tk = \infty} \nn \\
=& \frac{2}{1 + 1 / \nu}\left[\frac{1 - i}{2\sqrt{2}}\sqrt{\frac{\pi}{(1 + 1 / \nu) \tilde{t}}} - \tilde{\Lambda}\right].
\label{eq.high_momentum_contribution}
\end{align}
Here we expand the primitive at the lower limit $\tk = \tilde{\Lambda}$ to lowest non-vanishing order in $\tilde{t}$. Adding the low and high momentum parts, neglecting the term of $\calO(\tilde{t})$, and with a bit of rewriting, we get
\begin{align} 
\Gamma_p(t) =& \frac{\sqrt{2}(1 + 1 / \nu)^2}{4\pi \tB}\frac{a^2}{\aB \xi}\int \! {\rm d}\cos \theta \int_0^{\tilde{\Lambda}} \!\! {\rm d}\tk \left[-\frac{\tk^3}{\sqrt{1 + \tk^2}} \frac{1}{\Phi_{\tp}(\tk)} + \frac{1}{1 + 1/\nu}\right] \nn \\
&- \frac{1}{4\tB}\sqrt{\frac{1 + 1 / \nu}{\pi}}\frac{a^2}{\aB \xi}\frac{1-i}{\sqrt{\tilde{t}}} = \Sigma_p - \frac{1}{4\tB}\sqrt{\frac{1 + 1 / \nu}{\pi}}\frac{a^2}{\aB \xi}(1 - i)\sqrt{\frac{\tB}{t}}.
\label{eq.short_time_Gamma}
\end{align}
$\Lambda$ now takes on the role of an initial momentum cut-off. The initial expansion, \eqref{eq.low_momentum_contribution}, only works for finite $\Lambda$, but the above expression is valid for $t \ll \tB$ for any large value of $\Lambda$. Importantly, we recognize the integral as the second order polaron energy shift, $\Sigma_p$, which is independent of $\Lambda$ as long as it is sufficiently high. As for the integral of $\Gamma_p$ we get to linear order in $t$

\begin{equation}
\int_0^{t} {\rm d}s \; \Gamma_p(s) \simeq (i - 1)\sqrt{\frac{t}{t_0}} + \Sigma_p t, 
\label{eq.intGamma_short_time}
\end{equation}
where we define $1 / t_0 = 16\pi (1 + 1/\nu) n_{\rm B}^2 a^4 / \mB = 16\pi n_{\rm B}^2 a^4 / m_r $. The coherence at short times is then
\begin{align}
C_p(t) &= C_p(0)\te^{-i(p^2 / 2m + \nB \T + \Sigma_p)t}\te^{i\int_0^t {\rm d}s \Gamma_p(s)} \nn \\
&\simeq C_p(0)\te^{-i(p^2 / 2m + \nB \T)t}\te^{-(1 + i)\sqrt{t / t_0}}, 
\label{eq.coherence_short_times_appendix}
\end{align}
which explicitly shows that the coherence does not depend on the polaron energy at short times, only the overall mean-field shift $\nB \T$ is present. Also, the decay of $|C_p(t)|$ is independent of momentum. In the ideal Bose gas limit, $\aB \to 0$, the coherence collapses to this short-time behaviour for all times. We show this explicitly in the zero momentum case in figure \ref{fig.Cnorm_aB_depend}. Here the approach to $|C_p(t)| = |C_p(0)|\exp(-\sqrt{t / t_0})$ is apparent. In this limit the polaronic properties, the energy shift $\Sigma_p$ and the residue $Z_p$, do not appear and the polaron cannot be defined at any stage of the dynamics. 

\section{Analytical expression for $\Gamma_0$ in the equal mass case} \label{app4}
We find a closed form expression for the time-dependent rate coefficient \eqref{eq.Gamma} for zero momentum and equal masses. The unitless form of the rate coefficient \eqref{eq.ImGamma2} becomes in the equal mass case $\nu = m / \mB = 1$
\begin{equation}
\Gamma_p(\tilde{t}) = -\frac{\sqrt{2}}{{\pi}\tB}\frac{a^2}{\aB\xi}\int {\rm d}\cos(\theta){\rm d}\tk \frac{\tk^3 \te^{-i\Phi_{\tp}(\tk)\tilde{t}}}{\sqrt{\tk^2 + 1} \Phi_{\tp}(\tk)}, 
\label{eq.Gammaunitless}
\end{equation}
with $\tilde{t} = t / \tB$. Here $\cos\theta \in (-1, 1)$ and $\tk \in (0, \infty)$. For zero impurity momentum $p = 0$ and equal masses, $\Phi$ has the simple inverse: $\tk = \frac{\Phi}{\sqrt{1 + 2\Phi}}$. Using this we get
\begin{equation}
\Gamma_0(\tilde{t}) = -\frac{2\sqrt{2}}{{\pi}\tB}\frac{a^2}{\aB\xi}\int_0^{\infty}{\rm d}\Phi \frac{\Phi^2}{(1 + 2\Phi)^{5/2}}\te^{-i\Phi\tilde{t}}. \nn
\end{equation}
The problem has thus been reduced to calculating the above Fourier transform. A primitive to the integrand is found in Mathematica. With a bit of rewriting we then arrive at the result
\begin{align}
\Gamma_0(\tilde{t}) =& -\frac{2\sqrt{2}}{{\pi}\tB}\frac{a^2}{\aB\xi}\left[\frac{(1 + i)\sqrt{\pi}}{24\sqrt{\tilde{t}}}\left(i\tilde{t}^2 + 6\tilde{t} - 3i\right)\te^{i\tilde{t}/2} \right. \nn \\
&\times \left.\left(1 - {\rm erf}\left(\frac{1 + i}{2}\sqrt{\tilde{t}}\right) \right) - \frac{5 + i\tilde{t}}{12} \right],
\label{eq.Gammaanalytical}
\end{align}
with ${\rm erf}$ the error function. This has also been checked numerically. It is also possible to achieve an analytical expression for the integral of $\Gamma_0(t)$. Integrating \eqref{eq.Gammaanalytical} we get
\begin{align}
\int_0^t {\rm d}s \; \Gamma_0(s) =& \frac{\sqrt{2}}{3\pi}\frac{a^2}{\aB \xi}\left[ \tilde{t} + i (i\tilde{t} + 3)\frac{1 + i}{2}\sqrt{\pi \tilde{t}}\;\te^{i\tilde{t}/2} \right. \nn \\
&\times \left.\left(1 - {\rm erf}\left(\frac{1 + i}{2}\sqrt{\tilde{t}}\right)\right) \right] \nn \\
&\simeq i\frac{2\sqrt{2}}{3\pi}\frac{a^2}{\aB\xi}\left(1 + \frac{3}{\tilde{t}^2}\right), 
\label{eq.Gammaintegralanalytical}
\end{align}
where the last expression is accurate \textit{only} asymptotically to order $\tilde{t}^{-2} = (\tB/t)^2$.

\section{Damping in the long time limit} \label{app5}
We calculate the polaron life-time, $1/(2\tau_p) = \lim_{t \to \infty} \Im \Gamma_p(t)$. We can write the damping as
\begin{equation}
\Im \Gamma_p(t) = \nB \T^2 \int\frac{{\rm d}^3k}{(2\pi)^3} \frac{\varepsilon^{\rm B}_\bk}{E_\bk} \Re \int_{0}^{t} {\rm d}s \; \te^{+i(\varepsilon_{\bp} - \varepsilon_{\bp - \bk} - E_{\bk}) s}. \nn
\end{equation}
For $t \!\! \to\!\! \infty$ the temporal integral yields $\int_{0}^{\infty} {\rm d}s \; \te^{+ixs} = {\rm Pr}(-i / x ) + \pi \delta(x)$, where Pr stands for the principal value and $x = \varepsilon_{\bp} - \varepsilon_{\bp - \bk} - E_{\bk}$. Taking the real part of this expression gives the damping rate at long times
\begin{equation}
\Im\Gamma_p(\infty) = \frac{1}{2\tau_p} = \nB \T^2 \int\frac{{\rm d}^3k}{(2\pi)^3} \frac{\varepsilon^{\rm B}_\bk}{E_\bk} \pi \delta(\varepsilon_{\bp} - \varepsilon_{\bp + \bk} - E_{\bk}), \nn
\end{equation}
with $\tau_p$ the polaron life-time. For momenta below the Landau critical momentum, $p < p_c = mc$, the integral is 0. For $p > p_c$ the integral can be evaluated analytically to yield
\begin{equation}
\frac{\tB}{2\tau_p} = \frac{(1 + 1 / \nu)^2}{4\sqrt{2}}\frac{a^2}{\aB\xi}\frac{\tk_{\rm max}\sqrt{1 + \tk_{\rm max}^2} - {\rm arcsinh}(\tk_{\rm max})}{p / p_c},
\label{eq.dampinglongtimelimit}
\end{equation}
for $p > p_c$, where we use $\tp_c = p_c \xi / \sqrt{2} = \nu / 2$ and
\begin{equation}
\tk_{\max} = \frac{\nu^2}{\nu^2 - 1}\left(\sqrt{\left(\frac{p}{p_c}\right)^2 + \frac{1}{\nu^2} - 1} - \frac{1}{\nu}\frac{p}{p_c}\right), 
\label{eq.kmaxlongtimedamping}
\end{equation}
with $\tk_{\max} = (p / p_c - p_c / p) / 2$ in the equal mass case $\nu = 1$. We hereby have a closed form expression for the long time damping for all mass ratios. 

\section{Momentum dependent residue} \label{app6}
\begin{figure}[t!]
\begin{center}
\includegraphics[width=0.55\columnwidth]{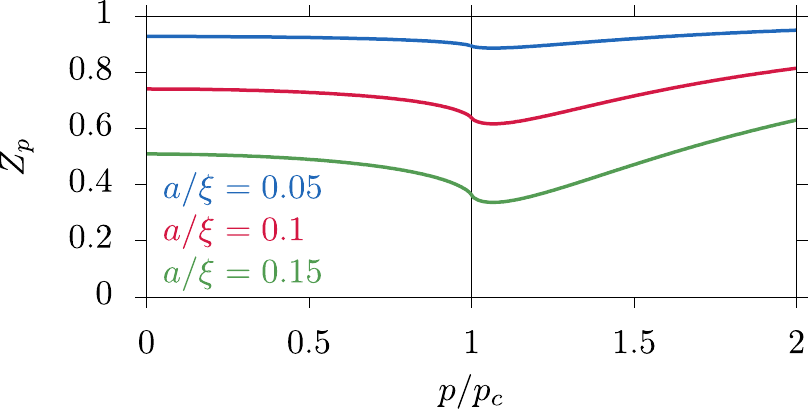}
\caption{Quasiparticle residue $Z_p$ as a function of the impurity momentum $p$ for different indicated values of the impurity-boson scattering length $a$ in the equal mass case $\nu = 1$. The functional behaviour is independent of the scattering lengths as evident from \eqref{eq.Zfinal} and exhibits a non-analytical kink at $p = p_c$ and a minimum value slightly above $p_c$. The boson-boson scattering length is $\aB / \xi = 0.01$.}
\label{fig.residue}
\end{center}
\end{figure}

We calculate the quasiparticle residue as given in \eqref{eq.residue}
\begin{equation}
Z_p = \te^{-\int_0^{\infty} {\rm d}t (\Im \Gamma_p(t) - 1 / 2\tau_p )}.
\label{eq.momentumdependentresidue}
\end{equation}
Using that $\pa \Phi_{\tp}(\tk) / \pa \cos(\theta) = -2\tp\tk / \nu$ we can perform the angle integral in \eqref{eq.Gammaunitless} and get
\begin{align}
\!\!\Im\Gamma_p(t) =& \frac{\sqrt{2}(1 + 1 / \nu)^2}{4\pi\tB}\frac{a^2}{\aB \xi}\frac{\nu}{2\tp} \int_0^{\infty} {\rm d}\tk \frac{\tk^2}{\sqrt{1 + \tk^2}} \left[{\rm Si}\left(\Phi^{+}_{\tp}(\tk)\tilde{t} \right) - {\rm Si}\left(\Phi^{-}_{\tp}(\tk)\tilde{t} \right) \right], \!\!
\label{eq.ImGammaintegral}
\end{align}
with $\tilde{t} = t / \tB$, $\tk = k\xi / \sqrt{2}$, $\nu = m / \mB$, ${\rm Si}(x) = \int_0^x {\rm d}u \sin(u)/u$ the sine integral, and $\Phi^{\pm}_{\tp}(\tk) = \tk \sqrt{\tk^2 + 1} + (\tk^2 \pm 2\tk\tp)/\nu$. Then integrating the sine integrals we get
\begin{align}
&\int_0^t {\rm d}s \; \Im\Gamma_p(s) = \frac{\sqrt{2}(1 + 1 / \nu)^2}{4\pi}\frac{a^2}{\aB \xi}\frac{\nu}{2\tp} \int_0^{\infty} {\rm d}\tk \frac{\tk^2}{\sqrt{1 + \tk^2}} \nn \\
&\times \left[ \tilde{t}\left({\rm Si}\left(\Phi^{+}_{\tp}(\tk)\tilde{t} \right) - {\rm Si}\left(\Phi^{-}_{\tp}(\tk)\tilde{t} \right) \right) + \frac{\cos\left(\Phi^{+}_{\tp}(\tk) \tilde{t}\right) - 1}{\Phi^{+}_{\tp}(\tk)} - \frac{\cos\left(\Phi^{-}_{\tp}(\tk) \tilde{t}\right) - 1}{\Phi^{-}_{\tp}(\tk)}\right]. \nn 
\end{align}
We recognize the first two terms in the integral as $\Im\Gamma_p(t) t$ which approaches $t / (2\tau_p)$ for $t \gg \tB$. We therefore get
\begin{align}
\ln Z_p =& -\int_0^{\infty} {\rm d}s \left[\Im\Gamma_p(s) - \frac{1}{2\tau_p} \right] \nn \\
=& -\lim_{\tilde{t} \to \infty} \frac{\sqrt{2}(1 + 1 / \nu)^2}{4\pi}\frac{a^2}{\aB \xi}\frac{\nu}{2\tp}\int_0^{\infty} {\rm d}\tk \frac{\tk^2}{\sqrt{1 + \tk^2}} \nn \\
&\times \left[\frac{\cos\left(\Phi^{+}_{\tp}(\tk) \tilde{t}\right) - 1}{\Phi^{+}_{\tp}(\tk)} - \frac{\cos\left(\Phi^{-}_{\tp}(\tk) \tilde{t}\right) - 1}{\Phi^{-}_{\tp}(\tk)}\right] \nn \\
=& -\frac{\sqrt{2}(1 + 1 / \nu)^2}{2\pi}\frac{a^2}{\aB \xi} \; {\rm Pr} \! \int_0^{\infty} \! \! {\rm d}\tk \frac{\tk}{\sqrt{1 + \tk^2}} \frac{1}{\left(\sqrt{1 + \tk^2} + \tk / \nu \right)^2 - \left(2\tp / \nu\right)^2}. 
\label{eq.Ainitial}
\end{align}
In the second equality we use that $(\cos(\Phi \tilde{t}) - 1) / \Phi = -\int_0^{\tilde{t}} {\rm d}\tilde{s} \sin(\Phi \tilde{t}) = \Im \int_0^{\tilde{t}} {\rm d}\tilde{s} \exp(-i\Phi \tilde{t}) \to {\rm Pr}(- 1 / \Phi) $, for $\tilde{t} \to \infty$, with Pr the principal value. The resulting integral above can readily be solved in the equal mass case $\nu = 1$. Here, using $x = \sqrt{1 + \tk^2} + \tk$ we get
\begin{align}
\ln Z_p &= -\frac{\sqrt{2}}{\pi}\frac{a^2}{\aB \xi} \; {\rm Pr}\int_1^{\infty} {\rm d}x \frac{x^2 - 1}{x^2(x^2 - (2\tp)^2)} \nn \\
&= -\frac{\sqrt{2}}{\pi}\frac{a^2}{\aB \xi}\frac{1}{(2\tp)^2}\left[1 + \frac{(2\tp)^2 - 1}{2\tp}\Re \; {\rm arctanh}\left(\frac{1}{2\tp} \right)\right]. \nn
\end{align}
The integral is performed using the transformation $\tanh(\theta) = x / 2\tp$. Note that $\Re \; {\rm arctanh}(1 / x) = \Re \; {\rm arctanh}(x) $. The final result is then
\begin{align}
\ln Z_p =& - \!\!\frac{a^2}{\aB \xi}\frac{\sqrt{2}}{\pi}\!\left(\frac{p_c}{p}\right)^{\!\!2}\!\!\left[\!1 + \frac{\left(\frac{p}{p_c}\right)^{\!2} - 1}{p / p_c}\Re \; {\rm arctanh}\left(\frac{p}{p_c}\right)\!\!\right] \nn \\
=& - \frac{a^2}{\aB \xi} \; A_p(1),
\label{eq.Zfinal}
\end{align}
thus defining $A_p(\nu)$ for $\nu = 1$. At $p = 0$ and $p = p_c$ a limiting process must be performed. We have thus obtained an analytical result for the residue as a function of momentum in the equal mass case. We plot the result for different values of the boson-impurity scattering length in figure \ref{fig.residue}. This exhibits an intriguing non-monotonic behaviour around $p = p_c$. 

Below we give the limiting values of $A_p(\nu)$ for general $\nu$ calculated from \eqref{eq.Ainitial} 
\begin{align}
A_0(\nu)	 	&= \frac{\sqrt{2}}{2\pi}\frac{\nu + 1}{\nu - 1}\left(1 - \frac{2}{\nu + 1} f(\nu)\right), \nn \\
A_{pc}(\nu) 	&= \frac{\sqrt{2}}{2\pi}\frac{\nu + 1}{\nu - 1}\ln\nu, 
\label{eq.Aalpha}
\end{align} 
where $f(\nu) = \sqrt{\frac{\nu + 1}{\nu - 1}} \; {\rm arctan}\sqrt{\frac{\nu - 1}{\nu + 1}}$ and $\sqrt{-1} = i$. Here $A_0(\nu = 1) = 2\sqrt{2}/(3\pi)$

\section{Asymptotic dynamics} \label{app7}
We calculate the asymptotic behaviour of the coherence for general mass ratios, $\nu$, and momenta, $p$. It turns out that there are two regimes in which this behaves quite differently: $p < p_c$ and $p > p_c$, where $p_c = mc$ is the Landau critical momentum of the condensate. \\

$\bf{p < p_c}$: We calculate $\int_0^t {\rm d}s\, \Gamma_p(s)$ to leading order in $1 / t$. We will see that
\begin{equation}
\int_0^t {\rm d}s \; \Gamma_p(s) \simeq \frac{a^2}{\aB \xi} \left( A_p(\nu) + B_p(\nu) \left(\frac{\tB}{t}\right)^2 \right), 
\label{eq.Gammapasymp1}
\end{equation}
where $A_p(\nu)$ is defined in \eqref{eq.Zfinal}. This analysis is detailed below. \\

Defining $w_{\tp}(\tk) = \tk^3 / (\sqrt{\tk^2 + 1}\Phi_{\tp}(\tk))$, we can write the time-dependent rate coefficient \eqref{eq.ImGamma2} as
\begin{equation}
\Gamma_p(t) = \! - \frac{\sqrt{2}(1 + 1 / \nu)^2}{4\pi \tB}\frac{a^2}{\aB \xi} \!\int {\rm d}\cos \theta {\rm d}\tk \; w_{\tp}(\tk) \te^{-i\Phi_{\tp}(\tk)\tilde{t}},
\label{eq.ImGamma3}
\end{equation}
We wish to expand this integral asymptotically for $\tilde{t} = t / \tB \gg 1$. Specifically we investigate the $\tk$-integral and write
\begin{equation}
I_{\theta} = \int {\rm d}\tk \; w_{\tp}(\tk) \te^{-i\Phi_{\tp}(\tk)\tilde{t}} = \frac{1}{-i \tilde{t}} \int {\rm d}\tk \frac{w_{\tp}(\tk)}{\pa_{\tk} \Phi_{\tp}}\; \pa_{\tk} \te^{-i\Phi_{\tp}(\tk)\tilde{t}}, \nn \\
\end{equation}
which is a simple rewritting valid as long as $\pa_{\tk}\Phi_{\tp} \neq 0$ for all $\tk$. This turns out to be the case as long as we are below the critical momentum, $p < p_c = mc$, or in units of $\xi$: $\tp = p\xi / \sqrt{2} < \nu / 2$. Then since the integral is now written as a function times the derivate of another we can use integration by parts. Repeating this procedure successively gives a systematic asymptotic expansion in powers of $1 / \tilde{t}$. It turns out that we have to go to order $1 / \tilde{t}^3$ to get a nonzero contribution
\begin{align}
I_{\theta} &= -\frac{i}{\tilde{t}^3}\!\left.\frac{\pa_{\tk}\left(\frac{\pa_{\tk} (w_{\tp} / \pa_{\tk}\Phi_{\tp})}{\pa_{\tk}\Phi_{\tp}}\right)}{\pa_{\tk}\Phi_{\tp}}\te^{-i\Phi_{\tp}(\tk) \tilde{t}}\right|_{\tk = 0}^{\tk = \infty} \!\!+ \mathcal{O}\left(\frac{1}{\tilde{t}^4}\right) = \frac{i}{t^3} \frac{2}{(1 - 2 \tp \cos \theta / \nu)^4} + \mathcal{O}\left(\frac{1}{\tilde{t}^4}\right). \nn
\end{align}
In the second equality we use that at low $\tk$ the functions asymptotically behave like $w_{\tp}(\tk) \to k^2 / (1 - 2\tp \cos \theta / \nu)$ and $\pa_{\tk}\Phi_{\tp} \to 1 - 2\tp \cos \theta / \nu$. 
Inserting this in \eqref{eq.ImGamma3}, and performing the angle integral we then get the asymptote
\begin{align}
\Im\Gamma_p(t)\tB \simeq& -\frac{1}{\tilde{t}^3} \frac{\sqrt{2}(1 + 1 / \nu)^2}{6\pi}\frac{a^2}{\aB \xi}\frac{\nu}{2\tp}\left(\frac{1}{(1 - 2 \tp / \nu)^3} - \frac{1}{(1 + 2 \tp / \nu)^3} \right). \nn
\end{align}
Integrating this and using $\int_0^{\infty} {\rm d}s\; \Im\Gamma_p(s) = A_p(\nu) \, a^2/(\aB \xi)$ we get
\begin{align}
&\int_0^t {\rm d}s \; \Im\Gamma_p(s) = \int_0^{\infty} {\rm d}s \; \Im\Gamma_p(s) - \int_{t}^{\infty} {\rm d}s \; \Im\Gamma_p(s) \nn \\
&\simeq \frac{a^2}{\aB \xi} \left[ A_p(\nu) + \left(\frac{\tB}{t}\right)^2 \frac{\sqrt{2}(1 + 1/\nu)^2}{12\pi}\frac{p_c}{p} \left(\frac{1}{(1 - p / p_c)^3} - \frac{1}{(1 + p / p_c)^3} \right) \right]. \nonumber
\end{align}
In the first line we use $\tilde t = t / \tB$, $\tp = p\xi / \sqrt{2}$, and $p_c\xi/\sqrt{2} = \nu/2$. Comparing with \eqref{eq.Gammapasymp1} we get
\begin{equation}
B_p(\nu) = \frac{\sqrt{2}(1 + 1/\nu)^2}{12\pi}\frac{p_c}{p}\left(\frac{1}{(1 - p / p_c)^3} - \frac{1}{(1 + p / p_c)^3} \right). 
\label{eq.Bp}
\end{equation}
Further, taking the $p = 0$ limit in the above, we get $B_0(\nu = 1) = 2\sqrt{2}/\pi$. This fits perfectly with the asymptotic form derived from an analytical expression for the integral in \eqref{eq.Gammaintegralanalytical}, which serves as a good check of our current expression. We have also successfully checked the asymptote numerically. This asymptotic expansion is accurate for $t / \tB \gg \sqrt{B_p / A_p}$. Evaluating $I_{\theta}$ to fourth order yields the dominant contribution to $\Re\Gamma_p$ at long times
\begin{align}
\Re\Gamma_p(t) \simeq & \left(\frac{\tB}{t}\right)^{\!\!4}\frac{3\sqrt{2}(1 + 1/\nu)^2}{2\pi \nu \tB}\frac{a^2}{\aB\xi}\frac{p_c}{p} \left(\frac{1}{(1 - p / p_c)^5} - \frac{1}{(1 + p / p_c)^5}\right).
\label{eq.ReGamma_asymp}
\end{align}

$\bf{p > p_c}$: A key feature for the above calculation is that the phase $\Phi_{\tp}(\tk)$ has no \textit{stationary} point, i.e. $\pa_{\tk}\Phi_{\tp} \neq 0$ for all $\tk$. For $p > p_c$ this is no longer true. As we shall now see, this leads to a different asymptotic behaviour of $\Gamma_p$. \\

The strategy in this case is slightly different. We take \eqref{eq.ImGammaintegral} and use the large argument asymptote of the sine integral ${\rm Si}(x) \simeq \pi / 2 - \cos (x) / x$
\begin{align}
\!\!\Im\Gamma_p(t)\tB \sim&\frac{\sqrt{2}(1 + 1 / \nu)^2}{4\pi}\frac{a^2}{\aB \xi}\frac{\nu}{2\tp} \int \! {\rm d}\tk \frac{\tk^2}{\sqrt{1 + \tk^2}} \!\!\left[\frac{\cos(\Phi^{-}_{\tp}(\tk)\tilde{t})}{\Phi^{-}_{\tp}(\tk)\tilde{t}} - \frac{\cos(\Phi^{+}_{\tp}(\tk)\tilde{t})}{\Phi^{+}_{\tp}(\tk)\tilde{t}} \right].\!\!
\label{eq.ImGammapasymp1}
\end{align} 
The phase $\Phi^{-}_{\tp}(\tk)$ has a stationary point $\tk_0$, i.e. $\pa_{\tk}\Phi^{-}_{\tp}|_{\tk = \tk0} = 0$, defined by the equation

\begin{equation}
0 = \sqrt{\tk_0^2 + 1} + \tk_0^2/\sqrt{\tk_0^2 + 1} + \frac{2}{\nu}(\tk_0 - \tp). 
\end{equation}

At the end of this section we will calculate $\tk_0$ in the limit $p \gtrsim p_c$. In general it turns out however, that this equation is quite involved to solve. Because of this stationary point the term in \eqref{eq.ImGammapasymp1} corresponding to $\Phi^{-}$ contains the leading order contribution at long times. We let $w^{-}_{\tp}(\tk) = \tk^2 / (\sqrt{1 + \tk^2}\Phi^{-}_{\tp}(\tk))$ and calculate the integral
\begin{align}
I_- &= \int_0^{\infty} {\rm d}\tk \; w^{-}_{\tp}(\tk) \te^{-i\Phi^{-}_{\tp}(\tk)\tilde{t}} \simeq \te^{-i\Phi^{-}_{\tp}(\tk_0)\tilde{t}} \int_{-\infty}^{\infty} {\rm d}\tk \; w^{-}_{\tp}(\tk) \te^{-i\Phi^{\prime \prime}_{\tp} / 2 \; (k - k_0)^2\tilde{t}} \nn \\
&\simeq \te^{-i\Phi^{-}_{\tp}(\tk_0)\tilde{t}}w^{-}_{\tp}(\tk_0) \sqrt{\frac{2\pi}{\Phi^{\prime \prime}_p\tilde{t}}} \te^{i\pi / 4}. \nn
\end{align}
First, we expand the phase to second order in $\tk - \tk_0$: $\Phi^{-}_{\tp}(\tk) \simeq \Phi^{-}_{\tp}(\tk_0) + \Phi^{\prime \prime}_p (k - k_0)^2 / 2 $. This is valid to do at long times, because away from $\tk_0$ the heavy oscillations lead to rapid cancellation. We also use this to expand the integral to $-\infty$. We use the rapid cancellation simply to evaluate the function in front at $\tk = \tk_0$. Finally, we evaluate the integral in the last step. Taking the real part of $I_-$ and dividing by $\tilde{t}$ we get the leading order contribution to the damping we were after
\begin{align}
\Im \Gamma_p \simeq& \frac{1}{2 \tau_p} + \frac{(1 + 1 / \nu)^2}{2\sqrt{\pi}}\frac{a^2}{\aB \xi} \frac{t_{\rm B}^{1/2}}{t^{3/2}}\frac{p_c}{p} \frac{w^{-}_{\tp}(\tk_0)}{ \sqrt{\Phi^{\prime \prime}_{\tp} } } \cos\left( \Phi^{-}_{\tp}(\tk_0)t / \tB + \pi / 4\right) \nn \\
					=& \frac{1}{2 \tau_p} + \frac{a^2}{\aB \xi} D_p(\nu) \frac{t_{\rm B}^{1/2}}{t^{3/2}} \cos\left( \omega_p t + \pi / 4\right). 
\end{align}
Here we have included the nonzero damping rate found in \eqref{eq.dampinglongtimelimit}, which the stationary phase approximation fails to calculate. We also define the expansion coefficient, $D_p$ and frequency of oscillation, $\omega_p$ according to
\begin{align}
D_p(\nu) &= \frac{(1 + 1 / \nu)^2}{2\sqrt{\pi}}\frac{p_c}{p} \frac{w^{-}_{\tp}(\tk_0)}{ \sqrt{\Phi^{\prime \prime}_{\tp} } }, \nn \\
\omega_p &= \Phi^{-}_{\tp}(\tk_0) / \tB.
\label{eq.Dp} 
\end{align}
Numerically, we have found that this expansion works very well after a few oscillations: $\omega_p t > 1$. Remember that $\pa_{\tk}\Phi^{-}_{\tp}|_{\tk = \tk_0} = 0$ and that $\bk_0$ is parallel to $\bp$ in $\Phi^{-}$. This explicitly shows that $\omega_p$ is the energy difference of the impurity before and after it has emitted a phonon in the forward direction. Further, this is at the minimum of the energy difference. For $p \to p_c^{+}$, $k_0 = \nu (p / p_c - 1)/2$ and in turn $\omega_p\tB = \nu (p / p_c - 1)^2 / 4$. 

\section{Divergence of formation time for a light impurity} \label{app8}
\begin{figure}[htb!]
\begin{center}
\includegraphics[width=0.55\columnwidth]{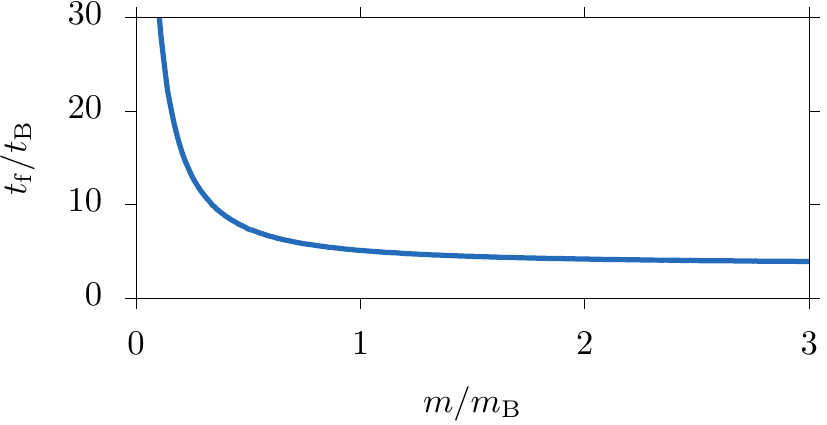}
\caption{The polaron formation time, $t_{\rm f}$, as a function of the mass ratio, $\nu = m / \mB$, in the zero momentum case. In these units $t_{\rm f}$ is independent of the scattering lengths $\aB$ and $a$, as along as we operate in the weakly interacting regime. The formation time diverges in the limit of a very light impurity, and quickly saturates to a value on the order of $\tB$ for a heavy impurity, $\nu > 1$. }
\label{fig.tf_alpha_dependence}
\end{center}
\end{figure}

We analyze the mass dependence of the polaron formation time, $t_{\rm f}$. This is defined and calculated as explained in Sec. \ref{sec.formationtime} of the main text. In figure \ref{fig.tf_alpha_dependence} we show the result for an impurity at zero momentum, $p = 0$. We clearly observe a divergence in the limit of a very light impurity, $\nu \to 0$. The underlying physical reason for this divergence is the enhanced recoil experienced by a light impurity. In terms of the dephasing picture developed in this article, see Sec. \ref{sec.formationtime}, the increasing formation time can be understood from the low momentum behaviour of the energy cost of generating phonons, $\Delta E_{\bp, \bk} = E_\bk + \varepsilon_{\bp - \bk} - \varepsilon_\bp$, which for zero impurity momentum, $p = 0$, and in units of $\tB$ reads
\begin{equation}
\Delta E_{p = 0, \bk} \cdot \tB = \tk \left(\sqrt{\tk^2 + 1} + \frac{\tk}{\nu}\right) \simeq \tk = \frac{\nu}{2}\frac{k}{p_c}, 
\end{equation}
valid for $k \ll p_c$. We hereby see, that the \textit{slope} of $\Delta E$ decreases with decreasing mass ratio $\nu = m / \mB$. Much like the slowdown around the critical momentum studied in Sec. \ref{sec.formationtime}, this diminished slope means that the scattering events for low momenta continue to add in phase for a longer time, thereby extending the period of time where damping takes place and increasing the formation time.

\section*{References}

\bibliography{ref_polform}

\end{document}